\documentclass[aps,prb,twocolumn,showpacs,superscriptaddress,floatfix]{revtex4}
\usepackage{graphicx}
\usepackage{amssymb}
\usepackage{amsmath}
\usepackage[english]{babel}
\usepackage{color}
\newcommand{\p}{{\phantom -}}  
\newcommand{\ostar}{\mbox{{$\otimes$}\llap{$\oplus$}}}  

\begin{document}
\title{Renormalization-group potential for quantum Hall effects}
\author{J. Nissinen}
\email{jaakko@fys.uio.no}
\author{C.A.  L\"utken}
\affiliation{Theory group, Department of Physics, University of Oslo, NO-0316 Oslo, Norway}
\date{\today}

\begin{abstract}
The phenomenological analysis of fully spin-polarized quantum Hall systems, based on holomorphic modular symmetries of the renormalization group (RG) flow, is generalized to more complicated situations where the spin or other ``flavors" of charge carriers are relevant, and where the symmetry is different.
We make the simplest possible ansatz for a family of RG potentials that can interpolate between 
these symmetries.  It is parametrized by a single number $a$  and we show that this suffices 
to account for almost all scaling data obtained to date. 
The potential is always symmetric under the main congruence group 
at level two, and when $a$ takes certain values this symmetry is enhanced to one of the maximal subgroups of the modular group. We compute the covariant RG $\beta$-function, 
which is a holomorphic vector field 
derived from the potential, and compare the geometry of this gradient flow with available 
temperature driven scaling data. 
The value of $a$ is determined from experiment by finding the location of a quantum critical point, i.e., an unstable zero of the $\beta$-function given by a saddle point of the RG potential. The data are consistent with $a \in \mathbb{R}$, which together with the symmetry leads to a generalized semi-circle law.
\end{abstract}

\pacs{73.43.-f, 73.20.-r}
\maketitle

\section{Introduction}
\noindent
The low energy physics that emerges at large scales from strongly interacting electrons in a background magnetic field and confined to two dirty dimensions, i.e., the quantum Hall effect (QHE), is not accessible by the conventional perturbative expansion of the local microscopic theory. Similar problems arise in high energy physics, where self-interacting gauge bosons give rise to an insoluble set of highly non-linear differential equations. One of the earliest and most successful methods developed to circumvent this problem used to be called ``phenomenological" \cite{Weinberg}, a precursor of what today is called effective field theory (which includes all quantum field theories in need of an ultraviolet cutoff). The essence of this idea is to use some of the global 
properties of the theory --- in the case of chiral models (i.e., low energy QCD and current algebra) the geometry of the global symmetries that are observed in the hadronic spectrum. These are always approximate and may be broken 
in some phases, giving rise to pseudo-Goldstone bosons (the pions) rather than massless bosons.

The ``phenomenological" approach to the QHE is similar in spirit, since it exploits emergent symmetries to harness the phenomenology of the system, but the symmetries are very different.  Since these are found to be infinite, discrete, non-abelian and holomorphic, a new set of mathematical tools is required, as well as a profoundly geometric understanding of the
renormalization group (RG) that we now describe.

\subsection{Renormalization}
\noindent
An RG flow is a vector field on the space of those parameters that are relevant at the chosen energy scale.  The QHE is parametrized by the conductivity tensor $\sigma^{ij}$, or equivalently its inverse, the resistivity tensor $\rho^{ij} = (\sigma^{-1})^{ij}$. 
These transport tensors are non-trivial because the background magnetic field breaks parity (time-reversal) invariance, whence the off-diagonal Hall coefficient is permitted by the generalized Onsager relation. A conventional Hall bar exhibiting fractional plateaux is usually a high mobility 
($\mu \sim 10^3- 10^6 cm^2/Vs$), low density ($n \sim 10^9 - 10^{11} /cm^2$) GaAs/AlGaAs heterostructure that confines all charge transport to a single isotropic layer of size $L_x\times L_y$, aligned with a current $I$ in the $x$-direction so that $R_{*x} = (L_*/L_y)\rho^{*x}$. The Hall resistance $R^\p_H = \rho^\p_H = \rho^{yx} = -\rho^{xy}$ is quantized in the fundamental unit of resistance, $h/e^2 = 25.812807557(18)\,k\Omega$, while the dissipative resistance 
$R^\p_D = \rho^{xx}/\Box$ is rescaled by the aspect ratio $\Box = L_y/L_x$.

We choose to label the low-energy parameter space by the complexified resistivity 
$\rho = \rho^{xy} + i\rho^{xx} = -\rho^\p_H + i\rho^\p_D$, 
or equivalently by the complexified conductivity 
$\sigma = \sigma^{xy} + i\sigma^{xx} = \sigma^\p_H + i\sigma^\p_D$. 
These complex coordinates take values in the upper half of the complex plane, 
$\mathbb H(\sigma) = \{\sigma \in \mathbb C\vert \Im\sigma = \sigma^\p_D > 0\}$, 
because the dissipative conductivity (resistivity) is positive. 
The reason for not including the real line in our definition of the parameter space 
will soon become clear.

The tangent vectors $\beta^1$ and  $\beta^2$ to the RG flow on the two dimensional 
space of conductivities are the physical (Gellman-Low) $\beta$-functions, 
which measure how fast the parameters (couplings) renormalize when the scale parameter 
$\Lambda$ changes:
\begin{equation*}
\beta^1= \frac{d\sigma^\p_H}{dt}, \quad 
\beta^2=\frac{d\sigma^\p_D}{dt}, \quad  t=\ln(\Lambda/\Lambda_0).
\end{equation*}

Very little can be said in general about the properties of RG flows, except that the topology of the 
flow is determined by the fixed points.
The flow ends at infra-red (IR) stable fixed points ($\oplus$), which in the quantum Hall case 
are the plateaux observed at rational values  $\sigma_\oplus = \sigma^\p_H \in \mathbb Q$ 
of the complexified conductivity.  This set $\mathbb P\subset \mathbb Q$ 
of IR fixed points are therefore the  \emph{only} real values that should be included in the 
physical parameter space, $\mathbb H_\oplus = \mathbb{H}\cup \mathbb P$.
Furthermore, physical (contravariant) $\beta$-functions have simple zeros at 
quantum critical points $\sigma_{\otimes}\in\mathbb{H}_\oplus$ 
for the localization-delocalization transition. 
Every member of this set $\mathbb E\subset \mathbb H_\oplus$ 
of critical points must be a proper saddle point of the flow, 
i.e., there should be both attractive and repulsive directions.

In Ref.~\onlinecite{LR1} it was proposed that a lot more can be said about RG flows in the QHE.
Careful examination of the fixed point structure (i.e., the set $\mathbb P\cup \mathbb E$ 
of stable and semi-stable fixed points) probed by scaling experiments reveal that there 
appear to be \emph{emergent symmetry} groups $\Gamma$ acting on the parameter space 
by fractional linear transformations $\gamma(\sigma) = (a\sigma + b)/(c\sigma + d)$.
The group elements $\gamma\in\Gamma$ are given by integers $a$, $b$, $c$ and $d$ 
satisfying $\det\gamma = \det(a,b;c,d) = ad - bc = 1$, and some additional constraints that distinguishes the different sub-groups of the full modular group ${\rm PSL}(2,\mathbb Z)$.
These so-called \emph{modular} symmetries can be used to constrain the RG flow, 
rendering them essentially unique in maximally symmetric cases.

In the absence of a rigorous derivation of the emergent symmetry $\Gamma$  from indisputable microphysics we have no other guide than experiment to aid us in the identification of $\Gamma$.  The full modular group is the simplest conceivable emergent symmetry,  but it is not of direct physical interest because it does not agree with data, as we shall discuss at length below.  
It is, however, the starting point for the mathematical discussion of these symmetries.
Our presentation mirrors the conventional one found in mathematics, which systematically develops the theory of modular transformations from the group theory of ${\rm PSL}(2,\mathbb Z)$ and its subgroups.  This may seem unnatural from a physics point of view, since ${\rm PSL}(2,\mathbb Z)$ is not a physical symmetry, but appears to be the only way to obtain a clear picture of these symmetries and how they are related. This is crucial in our work, since we are looking for a framework that interpolates between the different effective theories that are relevant for systems with different low energy degrees of freedom.

\subsection{Modular symmetry}
\noindent
This type of symmetry is not without precedent in condensed matter physics.
Kramers-Wannier duality is a discrete $\mathbb Z_2$-symmetry $D_{KW}$ 
that acts on the parameter space of the classical Ising model. It swaps the high temperature 
phase with the low temperature phase, which means that it is its own inverse, 
$D_{KW} = D_{KW}^{-1}$ or $D_{KW}^2 = 1$.  This is the prototype of the 
duality transformation that makes modular symmetries interesting.

The first and most important observation \cite{LR1} is that any $\Gamma$-symmetry partitions the 
parameter space into universality classes, 
with each phase ``attached" to a unique (plateau) fixed point $\oplus$ on the real line. 
This follows from the mathematical fact that compactifying the topology of the space on which a
modular group acts gives precisely the physical parameter space $\mathbb H_\oplus$.
\emph{Hall quantization is therefore an automatic and unavoidable consequence of modular symmetry.}  
This is the first example of a remarkable confluence of quantum Hall physics and modular mathematics.

\begin{figure}[t]   
\begin{center}
\includegraphics[scale = .6]{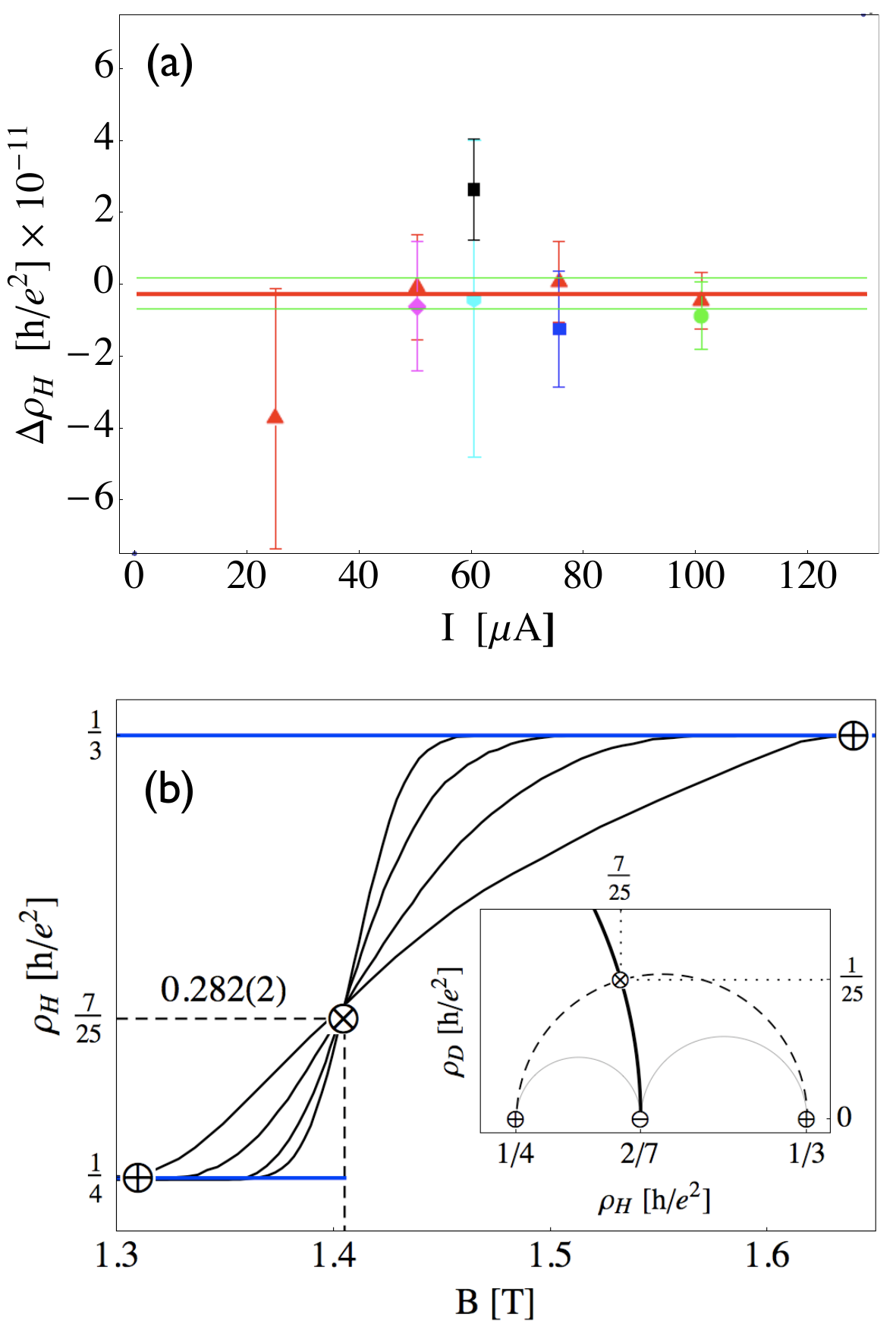}
\end{center}
\caption[Universality]{(Color online) Two experimental tests of the universal data encoded in a modular symmetry.
(a)~Measurements of $\Delta\rho_H^\p = \rho_H^\p ({\rm GaAs/GaAlAs})  - \rho_H^\p ({\rm Graphene})$ 
for the second Hall plateau $\rho_\oplus^\p = \rho_H^\p = 1/2\; [h/e^2]$ has verified
universality at the \emph{per trillion} level (adapted from Ref.\,\onlinecite{Janssen}). 
(b)~The location of the $3\otimes4$ quantum critical point, predicted 
by modular symmetry to be at $\rho_\otimes^\p = (7 + i)/25\; [h/e^2]$ (the relevant part 
of the modular phase diagram is shown in the inset) \cite{LR1}, 
has been verified experimentally at the \emph{per mille} level (adapted from Ref.\,\onlinecite{Tsui1}).}
\label{fig:Fig1}
\end{figure}

These quantum symmetries only emerge at temperatures that are so low, compared to 
the scale of the effective low energy modes of the quantum condensate, that thermal
fluctuations are swamped by quantum fluctuations.  So, unlike space-time 
or gauge symmetries, these global discrete parameter space 
symmetries are not exact, and we need to know how approximate they are.
The appropriate theoretical framework for addressing this question is effective field theory (EFT),
which is a systematic expansion in the inverse mass scale that separates the
microscopic physics (here quantum electro-dynamics in a disordered medium in a
strong background field) from the large scale physics being probed by transport 
experiments.  It is presumably the leading term in this expansion that is $\Gamma$-invariant.
Higher order terms may break the symmetry, but are suppressed by inverse powers 
of the mass gap.   If $\Gamma$ were exact the locations of all fixed points 
would be given by rational numbers, so the proximity of the experimental values to the
predicted rational values is a very sensitive test of the symmetry.
Two experiments of this type are shown in Fig.\,\ref{fig:Fig1}. 

Fig.\,\ref{fig:Fig1}(a) shows that \emph{universality}, which is the key concept powering 
the RG approach to EFT, has been verified in the QHE to an unprecedented accuracy \cite{Janssen}.
Any non-universal contribution to the plateau value was found to be less than a few parts 
\emph{per trillion}.
This is the reason that the QHE will be used as one of the corner stones in the new
SI system of metrology under construction.  These experiments have also found that 
the plateaux values are rational at the \emph{per billion} level.

If we interpret this as experimental evidence for a modular symmetry, 
then other consequences of modular symmetries should also be very accurate 
and provide further tests of the modular model.
For example, the fixed point set of each $\Gamma$ cannot be manipulated: 
all the plateaux and quantum critical points following from $\Gamma$ 
must be included \cite{LR1}.
As soon as one quantum critical point is pinned down the location of all the others 
is fixed by the symmetry.
In maximally symmetric cases, which includes the spin polarized system, 
there is no freedom at all: the exact location of all fixed points follow directly from the symmetry.
Only recently have scaling experiments probed sufficiently low temperatures for 
this idea to be properly tested, even in the simplest case of
 the fully spin polarized QHE  \cite{LR2}.

Fig.\,\ref{fig:Fig1}(b) compares an experiment \cite{Tsui1} determining the location of the 
quantum critical point in the delocalization transition between the the third and fourth 
Hall plateaux  ($3\otimes_\rho 4$ in our notation), with the location predicted by the
emergent modular symmetry identified two decades ago \cite{LR1, LR2}.
With the lowest temperatures around $10\,mK$ we expect that any 
modular symmetry is fully emerged and controlling all universal aspects of transport activity.
The agreement at the \emph{per mille} level inspires further confidence in the modular idea.
Observe that the only trace that appears slightly displaced from the theoretical value
$\rho_\otimes$ is obtained at the highest temperature ($\approx 0.5\,K$), 
and that it exhibits a very poorly articulated Hall quantization. This suggests that the system 
is not deeply into the quantum domain when the temperature exceeds a few hundred 
milli-Kelvin, and therefore is in a regime where the modular symmetry is expected to fail.
It is therefore rather surprising how accurate it still appears to be.  
Given a proper understanding of the separation of scales in this system,
it would be possible to estimate how accurate the EFT is, 
and therefore to quantify how approximate the symmetry is as a function of scale.

\subsection{Modular RG flow}
\noindent
Since the actions of the renormalization group and the modular group must be consistent, 
$\Gamma$-symmetry forces the RG flow into a straight-jacket \cite{LR1,BL1}. 
Not only are the RG fixed points, 
including the quantum critical (saddle) points, mapped into each other, so are the $\beta$-functions.
This means that the flow rates everywhere in one phase are mapped into the flow rates in any other phase,
and the flows observed experimentally in the spin-polarized QHE appear to satisfy this prediction 
as well \cite{LR2}.  In particular, it follows that the critical exponents $\lambda_\pm$, 
which are  the (inverse) flow rates in the principal directions near a critical point, 
are ``super-universal": they are always the same, independent of which 
quantum phase transition is considered \cite{LR1, BL1}.  This is consistent with 
measurements of the relevant exponent $(\lambda_+ > 0)$ for different Hall transitions
\cite{nu-exp}, as well as numerical ``experiments" \cite{nu-num}.
The irrelevant exponent $(\lambda_- < 0)$ appears to be all but impossible to measure, 
but numerical work reveals the striking possibility that 
$\lambda_- = - \lambda_+$ \cite{LR3}. 

This is evidence of an analytic structure that has been used to identify the exact geometry of the
RG flow \cite{BL1, LR4}.  We can combine the two real $\beta$-functions into one complex
function $\beta^\sigma = \beta^1 + i \beta^2$, but this $\beta$-function can not be holomorphic 
since holomorphic (complex analytic) functions do not have proper saddles, as required by physics.
This is seen by expanding a holomorphic function near a (simple) vanishing point $z_0$.  
There is only one real eigenvalue $\lambda = \lambda_+ = \lambda_-$, 
which cannot vanish since $z_0$ would then not be a simple zero.  
Consequently,  $z_0$ can be a source ($\lambda > 0$) or a 
sink ($\lambda < 0$) for the flow, but not a saddle point. \cite{BL1}
 
In another example of the convergence of quantum Hall physics and modular mathematics,
this is not an additional constraint, since holomorphic modular contravariant (physical)
vector fields do not exist at all. 
This follows from the observation \cite{BL1} that $\beta^\sigma$ transforms 
as a contravariant vector field when $\sigma\rightarrow \sigma^\prime = \gamma(\sigma)$ 
under a modular transformation $\gamma\in\Gamma$:
\begin{equation*}
\beta^{\sigma^\prime} =\left(\frac{d\sigma^\prime}{d\sigma}\right) \beta^\sigma
= (c\sigma + d)^{-2} \beta^\sigma\;,
\end{equation*}
where the modularity constraint $ad - bc = 1$ has been used.
A non-singular holomorphic function that transforms like this is called a modular form of weight $w = -2$, 
and the most basic fact of modular mathematics is that no form with negative weight exists.  
In other words, physics and mathematics are in agreement that such functions should not exist. 

The spectrum of holomorphic modular forms is very sensitive to the choice of symmetry group 
$\Gamma$.
If we want the form to be covariant under the full modular group ${\rm PSL}(2,\mathbb Z)$,
 then the lowest possible weight is $w = 4$.  
For the sub-groups found to be of relevance to the QHE the lowest weight is $w = 2$.
Since the covariant $\beta$-function has this weight,
\begin{equation*}
\beta_{\sigma^\prime} =\left(\frac{d\sigma}{d\sigma^\prime}\right) \beta_\sigma
= (c\sigma + d)^{+2} \beta_\sigma\;,
\end{equation*}
$\beta_\sigma$ can be both \emph{modular and holomorphic}.\cite{BL1}
The combination of these constraints gives an extremely rigid 
structure to the EFT, without which we would have no hope of 
proceeding along this path.  Real modular functions exist, 
but are all but impossible to work with.

If the EFT does have a complex structure respected by an emergent 
modular symmetry, then we are in unprecedented circumstances that 
allow the exact determination of global properties of the RG flow.
There are two independent arguments favoring this,
one experimental and one theoretical.

\subsection{Holomorphic RG flow}
\noindent
Observe first that a physical parameter space should be an ordinary 
Riemannian manifold with metric $G$,
which in our context must also be $\Gamma$-symmetric, i.e., 
a real modular form of weight $(w, \overline w) = (2,2)$.  
The natural geometry of $\mathbb H_\oplus$ is hyperbolic and the 
canonical metric is the Poincar\'e metric. This is the K\"ahler metric with components 
$\overline G_H = G_H = \partial_\sigma \partial_{\overline\sigma}K = 1/\sigma_D^2$, 
where $K = \ln(\sigma_D\vert f(\sigma)\vert^2)$ is a $\Gamma$-invariant 
K\"ahler potential and $f$ is, up to a phase (``multiplier system"), a holomorphic modular 
form of appropriate weight ($w = 1/2$). 
$K$ is a physically reasonable potential, in the sense described in this section, 
if $f$ is the Dedekind $\eta$-function, but we do not need detailed knowledge of the metric here.
Since we are assuming that a non-singular EFT exists for finite values of $\sigma$, 
this metric is non-singular at critical points and therefore invertible,
and we have:
\begin{equation}
\beta^{\rm phys} = \beta^{\sigma} = G^{\sigma\overline{\sigma}}\beta_{\overline{\sigma}}\;.
\label{eq:betafn}
\end{equation}
For finite values of $\sigma$ we can therefore quarantine non-holomorphicity of the physical $\beta$-function to the metric.
So-called ``holomorphic anomalies" may appear at singular values of $\sigma$, but they will not concern us here.
If $\beta_\sigma$ is holomorphic the pseudo-experimental fact that $\lambda_- = - \lambda_+$ 
follows from eq.\,(\ref{eq:betafn}), no matter what non-singular value the physical metric takes
at the critical points.

A second argument in  favor of a holomorphic $\beta_\sigma$ follows from our expectation
\cite{BL1} that
this vector field is a gradient flow. This means that
\begin{equation}
\beta_\sigma = - \partial_\sigma \Phi \;,
\label{eq:gradient}
\end{equation}
where  $ \partial_\sigma =  \partial/ \partial\sigma$ and the \emph{RG-potential} $\Phi\in\mathbb R$ is 
a kind of ``vacuum entropy" that counts the number of (effectively) massless degrees of freedom at critical points.

The existence of these potentials has been proven for two-dimensional unitary quantum field theories \cite{Zam} where they are known as ``C-functions". Quantum Hall dynamics is essentially two-dimensional since the phase space of incompressible quantum fluids is effectively two-dimensional (the spatial coordinates are canonically conjugate, as in Onsager's vortex dynamics). Furthermore, a similar result is expected to hold in any dimension,  so we may reasonably expect that the theorem applies.

This so-called ``C-theorem" guarantees that an RG potential exists that smoothly interpolates between conformal fixed points, which are critical points of the potential, and that the physical 
$\beta$-function is obtained from the gradient vector field generated by this potential using a metric on parameter space that will not concern us here. Suffice it say that it can be calculated directly from the EFT, if this is known. Since it is smooth and positive definite it does not affect the topology of the RG flow, only the absolute values of the flow rates. 
By construction this $\beta$-function is completely normal: 
it vanishes at critical points (where the tangents of the potential are flat), 
and the critical exponents are given by the principal curvatures of the potential at these saddle points.

While Zamolodchikov's proof is quite explicit, computation of his RG potential requires access to various correlations functions, i.e., essentially the full effective action. Since this is not available for the QHE a more oblique approach is needed. 
This is provided by Friedan's  proof of the C-theorem \cite{Garp}, which uses only general properties of spectral functions. This proof shows explicitly that the C-function  counts degrees of freedom, at least near the critical points where it equals the central charge of the conformal (scale-invariant) field theory. We recall some elementary properties of the spectral form. In two dimensions symmetries of space-time reduces the spectral function $\rho$ to a single scalar function, which by causality (unitarity) is positive definite. This function measures the density of degrees of freedom in the given theory, with poles at single-particle states and cuts at the continuum, which sum to unity. If the theory is changed so that a new set of low-energy states become relevant, then this theory has a different spectral function. We can imagine that these two theories belong to a single, one-parameter family of effective actions by a suitable choice of basis (lagrangian parameters or coupling constants), giving a  family of spectral functions $\rho_a\in\mathbb R$ ($a \in \mathbb R$) that interpolates smoothly between the two theories.
The corresponding family of RG potentials $\Phi_a$ is obtained by integrating the spectral function. 
Friedan's dissection of this potential shows that it must be a positive real function, 
and the C-theorem shows that the RG flow must be a gradient flow, i.e., ``downhill". 
As long as the effective theory is well behaved this should be true in any dimension 
since it avoids limit cycles and other pathologies of the flow. 
The existence of these potentials is a very useful simplification, since it lifts the geometric analysis of 
renormalization from vector to scalar fields, and RG potentials will therefore 
be the starting point for our generalized analysis presented below.

To this reasonable list of properties we add that the RG potential should be 
``holomorphically factorized", which means that:
\begin{equation}
\Phi(\sigma,\overline{\sigma}) = \ln\vert\varphi\vert^2 =  
\ln \varphi(\sigma) + \ln \overline{\varphi}(\overline{\sigma})\;. 
\label{eq:RGpotential}
\end{equation}
It is really the partition function 
$\mathcal Z(\sigma,\overline\sigma) = \vert\zeta\vert^2 = \zeta(\sigma) \overline\zeta(\overline\sigma)$
that factorizes into a holomorphic and anti-holomorphic part, but the terminology is inherited by the 
``thermodynamic" potentials derived from $\ln\mathcal Z$.   The vacuum energy, for example, is
 $F(\sigma,\overline\sigma) \propto\ln\mathcal Z(\sigma,\overline \sigma) = 
 \ln\zeta(\sigma) + \ln\overline\zeta(\overline\sigma)$,
 and we expect the ``vacuum entropy" $\Phi(\sigma,\overline\sigma)$ to ``factorize" in the same way.
This is automatically the case at quantum critical points (because of the local conformal symmetry), 
and it also agrees with the flows observed in spin-polarized quantum Hall systems
where a holomorphically factorized potential appears to account for all available scaling data.

It now follows from eq.\,(\ref{eq:gradient}) that the covariant $\beta$-function should be the
logarithmic derivative of a holomorphic potential $\varphi$:
\begin{equation}
\beta_\sigma = - \frac{\varphi^\prime(\sigma)}{\varphi(\sigma)}\;,\quad 
\beta_{\overline\sigma} = - \frac{\overline\varphi^\prime(\overline\sigma)}{\overline\varphi(\overline\sigma)}\;.
\label{eq:betafns}
\end{equation}
For a system with modular symmetry this implies that $\Phi_a$ is a real-valued modular function 
that is finite in the finite part of $\mathbb H$, and $\beta_\sigma$ is a holomorphic weight 2 modular form. 
A modular function is not, strictly speaking, a weightless modular form, since it must have a pole somewhere, 
but since we are studying the system at strong coupling it is physically reasonable to push this singularity to infinity, 
where the model decouples and we expect the $\sigma$-model to be irrelevant.
The existence of these potentials is automatic in modular mathematics, 
since weight $w = 2$ forms actually do derive from holomorphic potentials,
showing again how remarkably well suited modular mathematics is for the task at hand.

In short, we have argued that a \emph{physical RG potential} is given by a 
\emph{holomorphic modular function} $\varphi(\sigma)$, as in eq.\,(\ref{eq:RGpotential}).
While the argument was conceived in the context of the spin-polarized QHE, as presented here
there appears to be no obstruction, physical or mathematical, to considering more general situations.
It is therefore our purpose to test this conjecture in the widest available context.  In other words,   
emboldened by the success in the spin-polarized case, we seek other situations where the effective theory may be harnessed by an emergent modular symmetry. The first step is obviously to consider other quantum Hall experiments, where additional degrees of freedom are relevant. This includes situations where spin is not fully polarized, multi-layered systems, graphene, etc., collectively referred to as ``multi-component quantum Hall systems" in the following. 
In order to develop the vocabulary needed to discuss their symmetries it is advantageous to first summarize 
some basic facts about parameter space symmetries, modular transformations and RG flow geometry in the next section.
Equipped with the appropriate mathematics we then proceed to a detailed symmetry analysis 
of the observed phenomenology of some multi-component systems in Sect.\,3.

\section{Symmetries and RG potentials} 
\label{sec:potentials}
\noindent
Mathematically, the full modular group $\Gamma(1) = {\rm PSL}(2, \mathbb Z)$ 
is the simplest of the modular symmetries,  
but it does not agree with the scaling data in quantum Hall systems.
It is, however, the starting point for the conventional mathematical discussion of these symmetries,
reviewed here, which systematically develops the theory of modular transformations from the group theory of $\Gamma(1)$ and its subgroups.  This may seem unnatural from a physics point of view, since $\Gamma(1)$ is not a physical symmetry, but it gives a clear picture of how these symmetries are related. This is crucial in our work, since we are looking for a framework that interpolates between the different effective theories that are relevant for systems with different low energy degrees of freedom.  

We are fortunate that modular symmetries, i.e., subgroups $\Gamma\subset\Gamma(1)$,
appear prominently in many branches of mathematics, and therefore are extremely well studied.
Since there are infinitely many modular subgroups, it is also very fortunate that scaling experiments severely constrain the physically relevant symmetries.
As already mentioned $\Gamma$ cannot be too large, 
but it also cannot be too small, since it then does not have natural candidates 
for the quantum critical points of the plateau-transitions.
We can therefore confine attention to the most symmetric situations that yield 
phenomenologically viable RG flows, first considered systematically in the context of 
quantum phase transitions in Ref.\,\onlinecite{L1}. 
This includes only ``level 2" groups, which are the largest subgroups of  $\Gamma(1)$,
as we now explain.

\subsection{Duality}
\label{sec:duality}
\noindent
$\Gamma(1) = \langle T, S \rangle$ is generated by translations 
$T(\sigma ) = \sigma + 1$ and the simplest 
duality transformation $S(\sigma) = -1/\sigma$, which satisfy two (and only two) algebraic 
constraints: $S^2 = 1 = (ST)^3$.   They can be composed to give any integer value of the matrix elements in
$\gamma = (a,b;c,d)\in\Gamma(1)$.   This symmetry turns out to be too strong for physical applications, 
at least the ones considered here, so we consider the largest subgroups.  They are obtained either
by weakening the translation symmetry to $\;T^n(\sigma) = \sigma + n\;$ for some integer $n>1$,
or by relaxing the duality symmetry, or both.  

By a duality $D$ we mean a transformation analogous to 
Kramers-Wannier duality, so that  acting with $D$ twice we get back to the starting point ($D^2 = 1$).
Conjugating $S$ with any group element $X$ gives a transformation $D_X = X S X^{-1}$
that is self-dual because $S$ is ($S = S^{-1}$).
The other constraint, $(STS)(TST) = 1$,  suggests that $R = STS$ (the $S$-conjugate of $T$), 
or equivalently its inverse $W = R^{-1} = TST$, will figure prominently together with $S$ and $T$.
On the upper half plane $R^n(\sigma) = \sigma/(1 - n\sigma)$. 
The main congruence subgroup at level two is $\Gamma(2) = \langle T^2, R^2 \rangle$, 
which means that the matrix representation only contains matrices that reduce to the identity
mod 2 (i.e., $a, d = 1;\; b,c = 0 \mod 2$).
  
There are four groups of modular symmetries ``between" $\Gamma(1)$ and $\Gamma(2)$, 
i.e., sub-groups of $\Gamma(1)$ containing $\Gamma(2)$ as a sub-group, 
which are called ``level 2" symmetries.
The largest of these (it has index 2 in $\Gamma(1)$, 
see for example Ref.\,\onlinecite{Rankin} [p.5] for a definition) 
is $\Gamma_P = \Gamma_2 = \langle ST, TS \rangle$.
It is too large for our physical applications since it does not admit a physical potential.  
If it did the gradient of this potential would be a weight $w = 2$ modular form, 
and no such form exists for this group.  $\Gamma(1)$, which contains this 
sub-group, is eliminated from further consideration by the same argument.

If we want to keep all translations generated by $T$, then the group is unique:
$\Gamma_T = \langle T, D_R \rangle = \langle T, R^2 \rangle$.
This is essentially (up to outer automorphisms) the spin-polarized quantum Hall group.
Keeping instead the original duality $S$ requires that we double the translations to $T^2$, 
giving the group $\Gamma_S  = \langle T^2, S \rangle$.
Finally, the last of the groups at this level also has less translation symmetry,
$\Gamma_R = \langle R, D_T \rangle =  \langle T^2, R \rangle$.
These three \emph{maximal symmetries} are viable candidates for physical symmetries 
because they have physical potentials.   
We shall in fact see that these potentials belong to a $\Gamma(2)$-invariant
one-parameter family that interpolates between these points of enhanced symmetry.

\begin{figure}[t]   
\begin{center}
\includegraphics[scale = .9]{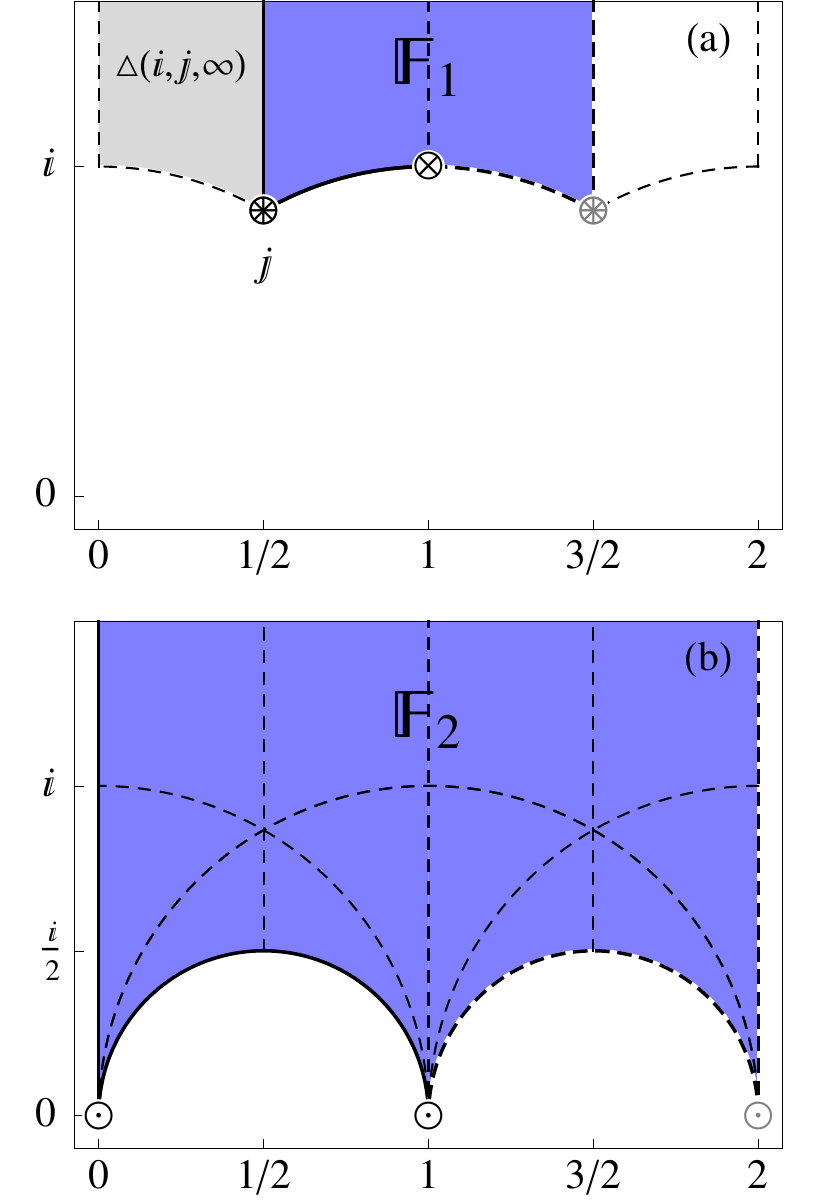}
\end{center}
\caption[FundamentalDomains12]{(Color online) Fundamental domains 
$\mathbb F_1$ and $\mathbb F_2$ for the main congruence groups 
$\Gamma(1) = {\rm PSL}(2,\mathbb Z)$ and  $\Gamma(2)$, 
showing conifold points of order two ($\otimes\in\mathbb H$) and three 
($\ostar\in\mathbb H$),  and the real cusps ($\odot\in\mathbb R$). Not shown is
the inequivalent cusp at infinity ($i\infty = \odot$).}
\label{fig:Fig2}
\end{figure}

If one of these symmetries emerges at low energy for a given physical situation, it is most easily recognized
by identifying the parities of the attractors on the real line, i.e., the parities of the plateaux values $\oplus = p/q$.
The modular group $\Gamma(1)$ does not distinguish between the parities of the fractions $p/q \in \mathbb Q$, so all rational
numbers are equivalent under this symmetry.  $\Gamma(2)$, on the other hand, respects the parities of 
both $p$ and $q$, so it partitions the rationals into three equivalence classes. Each of the index 3 groups partition the rationals 
into two equivalence classes. With ``$o$" representing odd integers and ``$e$" representing even integers:
\begin{align*}
\Gamma_T = \Gamma_0(2) = \langle T, R^2\rangle:  & \quad \{q \in e\} \cup \{q \in o\}\\
\Gamma_R = \Gamma^0(2) = \langle T^2, R\rangle: 
& \quad \{p \in e\} \cup \{p \in o\}\\
\Gamma_S = \Gamma_\theta(2)  = \langle T^2, S\rangle:  
& \quad \{pq \in e\} \cup \{pq \in o\}\; ,
\end{align*}
where we have included synonymous group names favored by mathematicians.

\begin{figure}[t]   
\begin{center}
\includegraphics[scale = 1.2]{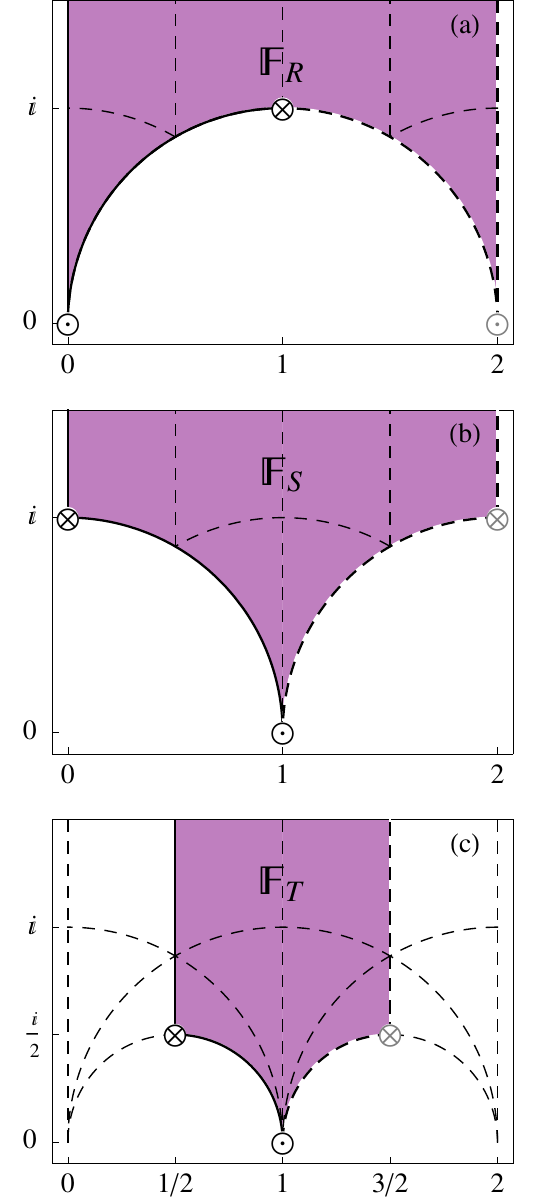}
\end{center}
\caption[FundamentalDomains]{(Color online) Fundamental domain 
(a) $\mathbb F_R$ for $\Gamma_R$ (b) $\mathbb F_S$ for $\Gamma_S$
 (c) $\mathbb F_T$ for $\Gamma_T$.  Conifold points $\otimes\in\mathbb H$ 
and real cusps $\odot$ are exhibited, the inequivalent cusp $i\infty = \odot$ is not.}
\label{fig:Fig3}
\end{figure}

A symmetry $\Gamma$ identifies points in $\mathbb H$, tesselating it with copies of the \emph{fundamental domain} $\mathbb F_{\Gamma}$, which can be chosen to be any
subset of $\mathbb H$ where distinct points cannot be connected with transformations 
in $\Gamma$.   The larger the symmetry the smaller the fundamental region is, and if 
$\Gamma$ has index $\mu$ in $\Gamma(1)$ then $\mathbb F_{\Gamma}$
contains $\mu$ copies of $\mathbb F_1 = \mathbb F_{\Gamma(1)}$.
 Fundamental domains $\mathbb F_1$ and $\mathbb F_2$ for the main congruence groups 
$\Gamma(1) = {\rm PSL}(2,\mathbb Z)$ and  $\Gamma(2)$
are shown in Fig.\,\ref{fig:Fig2}, and fundamental domains for congruence subgroups 
of index $\mu = 3$ are shown in Fig.\,\ref{fig:Fig3}.

The boundary of a fundamental domain may  contain \emph{conifold points}
where the symmetry has ``folded" a local disc into a conical shape.
When this occurs the fundamental domain is a conifold, i.e., 
a generalized manifold with conical singularities that look locally like cones. 
This happens at points that are fixed by some group element $X$,
and the deficit angle of the cone is $(1-1/n)\,2\pi$, where $n$ is  
the order of the group element, $X^n = 1$.

$\mathbb F_1$ has two such points.  Since $i = \sqrt{-1} = \otimes$ is fixed by 
the order 2 group element $S$ ($S^2 = 1$), the deficit angle is $\pi$. 
Similarly, since $j = \exp(2\pi i/3) = \ostar$ is fixed by the order 3 group element
$ST$ ($(ST)^3 = 1$), the deficit angle is in total $4\pi/3$.

We will always choose a fundamental domain that contains ``the point at infinity".
This point $i \infty$, and possibly some of its real images, are called \emph{cusps} ($\odot$), 
and by definition modular forms must not grow too fast near these points.
IR stable ($\oplus$) and UV unstable fixed points ($\ominus$) of the RG flow are cusps.

The pale critical points shown in Fig.\,\ref{fig:Fig2} are excluded from the fundamental 
domains to avoid double counting:
(a)~$j + 1 \simeq j = \ostar$ since $T\in\Gamma(1)$, 
(b)~$2 \simeq 0 = \odot$ since $T^2\in\Gamma(2)$.
The dashed parts of the boundary are excluded for the same reason.
Consider for example the solid semicircular boundary arc between $0$ and $1$ in 
 Fig.\,\ref{fig:Fig2}(b), given parametrically by $z(\theta) = (1 + \exp(i\theta))/2$
 with $\theta\in(0,\pi)$.  Using the product of the generators of $\Gamma(2)$
 this maps to $T^2R^2(z) = (2 - 3z)/(1-2z) = (3 -  \exp(-i\theta))/2$,
 which parametrizes the dashed semicircular arc between $1$ and $2$.
Since $z = 1$ is the fixed point of $T^2R^2$, the dashed arc is a mirror image of
the solid arc, both originating at this point.

Regions enclosed by dashed and solid lines are images of the hyperbolic triangle 
$\bigtriangleup(i,j,\infty)$.  For aesthetic reasons we have displayed 
$\mathbb F_1 = \bigtriangleup(i + 1,j,\infty) \cup \bigtriangleup(i+1,j+1,\infty)$,
rather than the more conventional choice 
$ \bigtriangleup(i,j-1,\infty) \cup \bigtriangleup(i,j,\infty)$.
$\mathbb F_2$ is six times as big (12 triangles) as $\mathbb F_1$ (2 triangles)
because $\Gamma(2)$ has index $\mu = 6$ in $\Gamma(1)$.
In the hyperbolic (Poincar\'{e})  metric all triangles have the same area, but  clearly not in the 
Euclidean metric used in these diagrams

Similarly, the dashed boundaries and pale critical points shown in Fig.\,\ref{fig:Fig3}
are excluded from the fundamental domain to avoid double counting:
(a)~$2 \simeq 0 = \odot$ since $T^2\in\Gamma_R$,
(b)~$i + 2  \simeq i = \otimes$ since $T^2\in\Gamma_S$,
(c)~$(3 + i)/2  \simeq (1 + i)/2 = \otimes$ since $T\in\Gamma_T$.
 $\mathbb F_R$\,, $\mathbb F_S$ and $\mathbb F_T$ 
are three times as big (6 triangles) as $\mathbb F_1$ because $\Gamma_R$\,, 
$\Gamma_S$ and $\Gamma_T$ have index $3$ in $\Gamma(1)$.

The subgroups considered here can only have conifold points of order 2. 
We shall see that $\beta$-functions must vanish at these points, 
but the converse is not true, i.e., $\beta$-functions can vanish at points 
that are not fixed by the group.
Only in maximally symmetric cases ($\mu = 3$) are RG and $\Gamma$ fixed points the same.
We use the icon $\otimes$ to represent these critical points, whether they are
fixed points of the group or not.  

Each symmetry $\Gamma\subset\Gamma(1)$  determines a unique pair of phase diagrams,
depending on whether the cuspoidal fixed point $i\infty \simeq 1/0 = \odot$ is attractive or repulsive.  
If $\Gamma$ acts on the parameter space of a model, 
the space of conductivities, say, then this is the repulsive UV 
fixed point  ($i\infty = \ominus$) around which a perturbative expansion is usually developed.
But if $\Gamma$  is acting on the inverse transport tensor (the resistivity), then it may be an
attractive IR fixed point ($i\infty = \oplus$).  Given this one bit of information $\Gamma$ 
determines the topology of the phase diagram.  This is particularly easy to see
in the maximally symmetric cases where $\Gamma = \Gamma_X$ $(X = R, S, T)$. 
The phase diagram is dictated by which phases and which phase transitions are possible.
If $i \infty$ is an attractive (repulsive) fixed point, then a transition
\begin{equation*}
f = p/q = \oplus \leftarrow\otimes \rightarrow\oplus^\prime = p^\prime/q^\prime = f^\prime>f
\end{equation*}
 exists for:
\begin{description}
\item[$\Gamma_{\rm T\phantom W}$] iff\;$q$ and $q^\prime$ are even (odd) and $\delta = 2 (1)$,
\item[$\Gamma_{\rm R\phantom T}$] iff \;$p$ and $p^\prime$ are odd (even) and $\delta = 1 (2)$,
\item[$\Gamma_{\rm S\phantom W}$] iff \;$pq$ and $p^\prime q^\prime$ are even (odd) and $\delta = 1 (2)$,
\end{description}
where $\delta = \det(f^\prime; f) = \det(p^\prime, p ; q^\prime, q) = p^\prime q - p q^\prime$.
These ``diagnostic rules" follow immediately from the parity properties of the group,
so the symmetry determines, and is determined by,  the fixed point structure.

Since the index 3 symmetry groups $\Gamma(2) \subset \Gamma_X\subset\Gamma(1)$
($X = R, S, T$) all have physical potentials they are the focus of our phenomenological analysis. 
These potentials belong to a $\Gamma(2)$-invariant one-parameter family, constructed next.

\subsection{Analytic structure}
\label{sec:analytic}
\noindent
We have so far concluded that we need to investigate the existence and uniqueness 
of meromorphic potentials and holomorphic vector fields that are automorphic 
under some modular symmetry.
The structure of a meromorphic function $f(z)$ is completely determined 
by the points where it is not finite, i.e., its zeros and poles.
By the order $n_p$ of a point $z_p$ in $\mathbb H$ we mean the 
leading (lowest) power of the Laurent expansion at that point,  i.e., 
 $f(z) \propto (z-z_p)^{n_p} +\cdots$.  If  $n_p = 0$ then $f(z_p)$ is finite ($\neq 0, \infty$)
 and $z_p$ is a regular point.   If  $n_p$ is positive (and finite) then $f(z_p)$ vanishes and 
 $z_p$ is a simple ($n_p = 1$) or multiple ($n_p > 1$) zero.
 If  $n_p$ is negative (and finite) then $f(z_p)$ is singular and 
 $z_p$ is a simple ($n_p = -1$) or multiple ($n_p < -1$) pole.
 If $n_p = \pm\infty$ the function is not meromorphic.
 
\begin{table*}[t]  
\begin{center}
\begin{tabular}{||c||c|c|c|c|c|c|c|c|c|c|c|c|c|c|c|c||c||}  \hline
&$a$&$\;w\;$&$\;\mu\;$&$\mu w/12$&$\sigma=0$&$n_0$&$\sigma=1$&$n_1$&
$\sigma=i\infty$&$n_\infty$&$\sigma=i$&$\;n_i\;$&$\sigma=j,j^\prime$&$n_{j,j^\prime}$&$\;n_B\;$&$\;n_F\;$&$\otimes\in\mathbb F$\\
\hline
$J$&$$&$0$&$1$&$0$&$\infty$&$-1\p$&$\infty$&$-1\p$&$\infty$&$-1\p$&$1$&$0$&$0$&$3$&$0$&$0$&$$\\ 
\hline
$\;\varphi_P^\p\;$&$$&$0$&$2$&$0$&$1$&$0$&$1$&$0$&$1$&$0$&$-1\p$&$0$&$0,\,\infty$&$\pm3\p$&$0$&$0$&$$\\ 
\hline
$\varphi_R^\p$&$-1\p$&$0$&$3$&$0$&$\infty$&$-2\p$&$0$&$1$&$0$&$1$&$2$&$0$&$-1\p$&$0$&$0$&$0$&$1+i$\\ 
\hline
$\varphi_S^\p$&$2$&$0$&$3$&$0$&$0$&$1$&$\infty$&$-2\p$&$0$&$1$&$-1/4$&$0$&$-1\p$&$0$&$0$&$0$&$i$\\ 
\hline
$\varphi_T^\p$&$1/2$&$0$&$3$&$0$&$0$&$1$&$0$&$1$&$\infty$&$-2\p$&$-2\p$&$0$&$1$&$0$&$0$&$0$&$(1+i)/2$\\ 
\hline
$\lambda$&$1$&$0$&$6$&$0$&$1$&$0$&$\infty$&$-1\p$&$0$&$1$&$\p1/2$&$0$&$-j,\,j^\prime$&$0$&$0$&$0$&$$\\ 
\hline
$\varphi_a^\p$&$a$&$0$&$6$&$0$&$0$&$$&$\infty$&$$&$0$&$$&$*$&$$&$*$&$$&$$&$$&$\sigma_\otimes(a)$\\ 
\hline
$\theta_2^4$&$-\infty$&$2$&$6$&$1$&$*$&$0$&$*$&$0$&$0$&$1$&$*$&$0$&$*$&$0$&$1$&$0$&$$\\ 
\hline
$\theta_3^4$&$0$&$2$&$6$&$1$&$*$&$0$&$0$&$1$&$1$&$0$&$*$&$0$&$*$&$0$&$1$&$0$&$$\\ 
\hline 
$\theta_4^4$&$1$&$2$&$6$&$1$&$0$&$1$&$*$&$0$&$1$&$0$&$*$&$0$&$*$&$0$&$1$&$0$&$$\\ 
\hline
$\beta_a$&$a$&$2$&$6$&$1$&$(1-a)*$&$0$&$a*\phantom{a}$&$0$&$1$&$0$&$*$&$0$&$*$&$0$&$0$&$1$&$\sigma_\otimes(a)$\\ 
\hline
$\beta_R^\p$&$-1\p$&$2$&$3$&$1/2$&$*$&$0$&$*$&$0$&$1$&$0$&$*$&$0$&$*$&$0$&$1/2$&$0$&$1+i$\\ 
\hline
$\beta_S^\p$&$2$&$2$&$3$&$1/2$&$*$&$0$&$*$&$0$&$1$&$0$&$0$&$1$&$*$&$0$&$1/2$&$0$&$i$\\ 
\hline
$\beta_T^\p$&$1/2$&$2$&$3$&$1/2$&$*$&$0$&$*$&$0$&$1$&$0$&$*$&$0$&$*$&$0$&$1/2$&$0$&$(1+i)/2$\\ 
\hline
\end{tabular}
\end{center}
\caption[longtablecapstion]{Analytic structure of some modular functions and forms at level $2$.
A star ($*$) represents a ``finite" but uninteresting value of the appropriate uniformizing variable.
 $\beta_a = i\pi (\theta_3^4 - a\theta_2^4)$ is ``physical" iff $a\neq0,1,\infty$, in which case it
 has a critical point  that is on $\partial\mathbb F$ if $a$ is real.  
 The last column contains the position of  the 
 saddle point of the potential, or equivalently, 
 the location of the simple zero of the $\beta$-function derived from this potential.
 $J$ has a double pole at $i\infty$ and triple zeros at both the conifold points 
 $j$ and $j^\prime = -j^2$, thus saturating the sum rule.
Similarly, $\varphi^{\phantom X}_P$ has a triple zero at 
$j$ and a triple pole at $j^\prime$, thus saturating the sum rule.}
\label{tab:Table1}
\end{table*}

For functions that transform with weight $w$  under a modular group 
the order of points on the boundary of the fundamental domain is not always obvious.
Consider first the cusp $i \infty$, which is the only cusp for $\Gamma(1)$.
We cannot Laurent expand $f(z)$ in $z - i\infty$, so by the order $n_{\infty}$ 
we mean in stead the leading power of an expansion in a suitable finite local
variable. Because level $N$ subgroups contain $T^N$, modular forms at level 
$N$ are periodic with period $2\pi N$, and can therefore be Fourier expanded 
in $q_N^\p = \exp(2\pi i /N)$. The order $n_{\infty}$ is the leading order in this 
$q_N^\p$-expansion.

For $N>1$ there are additional cusps on the 
real line that are images of the cusp at infinity.
The order $n_c$ of a weight $w$ function $f(z)$ at a cusp $z_c= \gamma_c(i \infty)$ 
is defined as the leading order of the $q_N^\p$-expansion of\break
$(cz+d)^{-w}f(\gamma_c(z))$.
For example, at level $N = 2$ the order $n_0$ of $f(z)$ at the origin $0 = S(i\infty)$ is the leading power of 
$z^{-w}f(-1/z)$ expanded in powers of $q_2 = \exp(i \pi z)$. 

Cauchy's theorem can be applied to a meromorphic function that transforms 
with weight $w$ under a subgroup of index $\mu$.   By choosing a contour along
the edge of $\mathbb F$, taking care to circumnavigate the cusps and conifold points 
counter clockwise and recalling that conifold points contribute less than a full arc,
it follows from Cauchy's theorem that \cite{Rankin}:
\begin{equation}
n_B^\p + n_F^\p  = \frac{\mu w}{12}\; ,
\label{eq:sumrule}
\end{equation}
where $n_{F}$ is the contribution from the finite part of $\mathbb H$.  $n_{B}$ is the contribution from special points (that are fixed by elements of the group) on the boundary of the fundamental domain,
\begin{equation*}
n_{B} = \sum_{c\in\partial\mathbb F} n_{c} + \frac{n_{\otimes}}{2} + \frac{n_{\ostar}}{3}\,,
\end{equation*} 
where $n_\otimes$ is the total order of all conifold points of order 2,
and  $n_{\ostar}$ is the total order of all conifold points of order 3
($\ostar$ contributes iff the symmetry is $\Gamma(1)$ or $\Gamma_{P}$). 
Any other singular points $p$ are in the finite part of $\mathbb F$, and 
\begin{equation*}
n_F^\p = \sum_{p\in\mathbb F}  n_p^\p
\end{equation*}
is the total order of these points.  The contribution from $n_F$ will typically be
severly constrained, or even eliminated, by the physical requirements of 
RG potentials and flows.

This sum rule is very useful since it allows us to classify the 
possible analytic behaviour of scalar and vector fields (potentials and $\beta$-functions),
i.e., the fixed points and singularities of effective field theories with $\Gamma$-symmetry.

For the full modular group $\Gamma(1)$, the contributions to the sum rule are
\begin{equation*}
n_{\infty} + \frac{n_{\otimes}}{2} + \frac{n_{\small\ostar}}{3} + n_F^\p = \frac{w}{12}.
\end{equation*}

Subgroups at level 2 of index 3 have two inequivalent cusps, one at $i\infty$ and 
another at either 0 or 1, and one conifold point of order two.  The sum rule is therefore 
\begin{equation*}
n_* + n_{\infty} + \frac{n_{\otimes}}{2}  + n_F^\p = \frac{w}{4} \quad  (* = 0,1)\; .
\end{equation*}
For low values of $w$ this sum rule is quite difficult to satisfy.
For example, since the $\beta$-function has weight 2
and the cusp orders and $n_F^\p$ are integers, we must have 
a critical point at $\otimes$ (i.e., a simple zero contributing $n_\otimes = 1$).  
Furthermore, since this function should derive from a physical potential
we are led to consider the simplest possible analytic structure that
saturates the sum rule, $(n_0, n_{\infty},  n_{\otimes}, n_F^\p) = (0, 0, 1,0)$.
 Remarkably, such entire and essentially unique $\beta$-functions 
 do exist, as we shall see below.

The only subgroup at level 2 of index 6 is the main congruence group $\Gamma(2)$,
which has three inequivalent cusps at $0$, $1$ and  $i\infty$, but no conifold points.
In this case the sum rule is 
\begin{equation*}
n_0 + n_1 + n_{\infty} + n_F^\p = \frac{w}{2} \; ,
\end{equation*}
which is a little easier to satisfy than the index 3 case.
For example, there are now two possible entire 
$\beta$-functions, which either has a simple zero
at one of the cusps and no other zeros, or a simple zero in the finite part of 
$\mathbb F$ and no other zeros.

\subsection{A potential family}
\label{sec:potential}
\noindent
The modular groups we have been discussing, 
$\Gamma(1),\, \Gamma_P,\, \Gamma_R,\, \Gamma_S,\, \Gamma_T,\,\Gamma(2)$, 
have canonical potentials  \cite{Rankin} (also called invariant or automorphic functions):  
\begin{eqnarray}
\varphi_1^\p &=&(\lambda^2 - \lambda + 1)^3  \lambda^{-2}\, (\lambda - 1)^{-2}  \nonumber\\ 
\varphi_P^\p &=& (\lambda + j)^3  (\lambda - j^\prime)^{-3}  \nonumber \\
\varphi_R^\p &=&  \;\;\;\lambda\, (\lambda - 1)^{-2} \nonumber \\  
\varphi_S^\p &=&  \;\;\;\lambda\, (\lambda - 1) \label{eq:potentials}\\
\varphi_T^\p &=&   \lambda^{-2}  (\lambda - 1)\nonumber \\
\varphi_2^\p &=&  \lambda = (\theta_2/\theta_3)^4 \nonumber\; ,
\end{eqnarray}
where
\begin{equation*}
j = e^{2\pi i/3} \;,\quad j^\prime = -j^2 = e^{\pi i/3} \; .
\end{equation*}
The analytic structure of these potentials, together with some other functions that will make an appearance
below, are listed in Tab.\,\ref{tab:Table1}.

The $\Gamma(1)$-invariant function $\varphi_1^\p$ is Klein's famous $J$-invariant,
 $J = (4/27)\, \varphi_1^\p$.
It has been expressed in terms of the $\Gamma(2)$-invariant function $\lambda$, 
which is the ratio of the Jacobi $\theta$-functions $\theta_2(\sigma) = \vartheta_2(0,q)$ and 
$\theta_3(\sigma) = \vartheta_3(0,q)$.  The elliptic $\vartheta$-functions $\vartheta_i$
are most conveniently given by their Fourier expansions:
\begin{eqnarray}
\vartheta_1(u,q) &=& 2 q^{1/4}\sum_{n=0}^{\infty} (-1)^n q^{n(n+1)} \sin((2n+1)u) \nonumber\\
\vartheta_2(u,q) &=& 2 q^{1/4}\sum_{n=0}^{\infty} q^{n(n+1)} \cos((2n+1)u) \nonumber\\
\vartheta_3(u,q) &=& 1 + 2\sum_{n=1}^{\infty} q^{n^2} \cos(2n u) \nonumber\\
\vartheta_4(u,q) &=& 1 + 2\sum_{n=1}^{\infty} (-1)^n q^{n^2} \cos(2n u) \; . \nonumber 
\end{eqnarray}
Since  $q = \exp(\pi i \sigma) = \exp(\pi i x) \exp(-\pi y)$, these sums converge very rapidly for reasonable values of $y =\Im\sigma$ and are therefore well suited for numerical work, especially graphics.

$J \propto \varphi^\p_1$ and  $\varphi^\p_P$ 
are too singular to give rise to a modular form of weight $w=2$.
There are in fact no weight 2 forms at all with the symmetries of these potentials,
so both $\Gamma(1)$ and its largest (index $\mu = 2$) subgroup $\Gamma_P$ are too 
large for physical applications and we need consider them no further.

The index three ($\mu = 3$)  subgroups are just barely small enough to admit
weight 2 forms, but only one each, with a single critical points (simple zero) that is forced to coincide with a fixed point
of the group.  An RG flow diagram with symmetry $\Gamma_R$, $\Gamma_S$ or $\Gamma_T$ is therefore extremely
rigid, predicting the exact location of all critical points.
The index six  ($\mu = 6$) potentials are more flexible, since there are two independent ones.  
They give rise to 
a 2-dimensional space of weight 2 forms, which we can choose to span with two of the Jacobi $\theta$-functions,
say $\theta_2^4$ and $\theta_3^4$.  Thus it follows from modular symmetry alone that $\theta_4^4$  
must be a linear combination of these, a fact discovered by Jacobi and recorded in his famous identity  $\theta_4^4 = \theta_3^4  - \theta_2^4$.

The purpose of all this mathematical machinery should now becoming clear.   
Of the six possible potentials at level two, $\varphi^\p_1$ and $\varphi^\p_P$ are unable to satisfy the 
physical constraints placed on their analytic structure. 
The simplicity and similarity of the remaining four potentials is striking, and suggests an obvious 
way to embed them into a single, one-parameter family of physical potentials.
Consider first the general $\Gamma(2)$-invariant RG potential 
$\Phi = \psi + \overline\psi$ obtained by a
linear superposition of the maximally symmetric potentials in eq.\,(\ref{eq:potentials}):
\begin{equation*}
\psi(\sigma) = c^\p_R \ln\varphi^\p_R +  c^\p_S \ln\varphi^\p_S + c^\p_T \ln\varphi^\p_T
\;,\quad c^\p_{R, S, T}\in\mathbb C.
\end{equation*}
This family interpolates smoothly between the maximally symmetric cases, 
but there are too many parameters for it too be useful as it stands \cite{L1}.  
We can ignore an arbitrary normalization factor 
that only affects flow rates, not their shapes, leaving two (complex) parameters.
However, since $\varphi^\p_R$,  $\varphi^\p_S$ and $\varphi^\p_T$ are simple rational functions 
in the fundamental $\Gamma(2)$ modular function $\lambda$, 
the level two symmetric family of quantum Hall RG potentials reduces to 
(we choose a convenient parametrization for $a$):
\begin{equation}
\Phi_a \propto \ln \lambda(\lambda-1)^{a-1} + c.c.
\label{eq:Phia}
\end{equation}

\subsection{Modular RG flows}
\label{sec:RGflow}
\noindent
From eq.\,(\ref{eq:betafns}) it immediately follows that the $\beta$-function
\begin{equation}
\beta_{a, \sigma} = - \partial_\sigma \Phi_a(\sigma) 
\propto  \frac{1 - a\lambda}{\lambda-1} \cdot \frac{\lambda^\prime}{\lambda}
= i\pi \theta_3^4 (1 - a\lambda)
\label{eq:betaa}
\end{equation}
by construction satisfies all physical requirements, thus proving the existence 
of a one-parameter family of physically viable $\beta$-functions that interpolates
between the enhanced symmetries $\Gamma_{R,T,S}$.
The second equality in eq.\,(\ref{eq:betaa}), which follows from the Jacobi identity and 
the fact that $\lambda' /\lambda = i\pi \theta^4_4$, is particularly useful for locating the
critical point. Since $\theta$-functions are finite on $\mathbb F$ (comp. Tab.\,\ref{tab:Table1}), 
$\beta$ can only vanish if 
$\lambda(\sigma_\otimes) = 1/a$, which can be inverted in terms of elliptic integrals 
of the first kind.  Consequently, the $\beta$-function has a simple zero at 
\begin{equation}
\sigma_\otimes = \lambda^{-1}(1/a) = \frac{i K'(1/a)}{K(1/a)} \mod \Gamma(2)\, ,
\label{eq:qcp}
\end{equation}
as it should, and nowhere else (in the fundamental domain), as it should.

\begin{figure}
\begin{centering}
\includegraphics[width=90 mm]{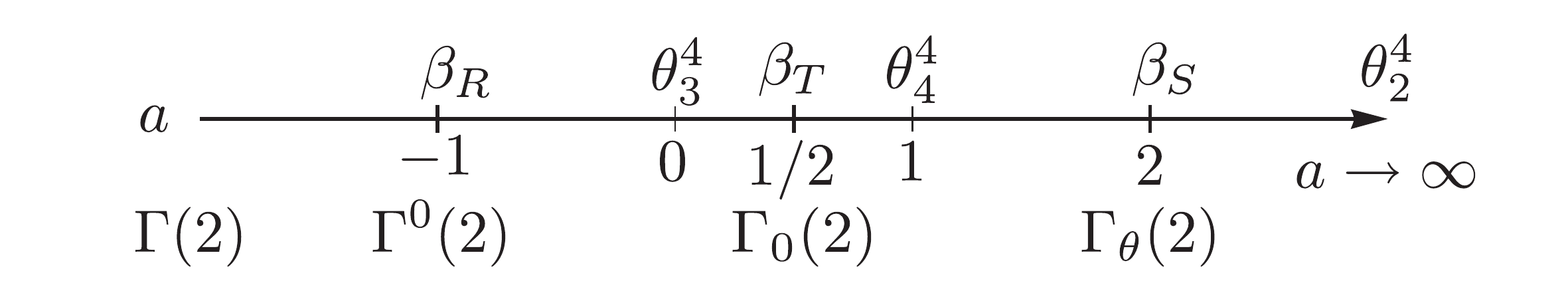}
\caption[summaryofavalues]{The flow of the vector field $\beta_a$ is $\Gamma(2)$ symmetric for any value of the parameter $a\in\mathbb C$.
When $a = -1, 1/2, 2$ the symmetry is enhanced to an index 3 group, and when $a = 0, 1, \infty$  
the $\beta$-function degenerates to a $\theta$-function, which is finite on the interior of the fundamental 
domain and therefore has no quantum critical point.}
\label{fig:Fig4}
\end{centering}
\end{figure}

The $\Gamma(2)$ symmetry of $\Phi_a$ is enhanced when $a$ takes certain values
(Fig.\,\ref{fig:Fig4}):
to $\Gamma_R$ when $a = a_R = -1$, 
to $\Gamma_T$ when $a = a_T = 1/2$ and 
to $\Gamma_S$ when $a = a_S = 2$.
The corresponding $\beta$-functions reduce to particularly simple Eisenstein functions \cite{BL1}:
\begin{eqnarray*}
\beta_R &\propto& \frac{\lambda + 1}{\lambda - 1} \cdot\frac{\lambda^\prime}{\lambda} \propto
1 + 24\sum_{n=0}^{\infty}  \frac{n {q}^n}{1 +  {q}^n}\\
\beta_S &\propto&  \frac{2\lambda - 1}{\lambda - 1} \cdot\frac{\lambda^\prime}{\lambda} \propto
1 - 24 \sum_{n=0}^{\infty}  \frac{(2n + 1) {q}^{2n + 1}}{1 +  {q}^{2n + 1}}\\
\beta_T &\propto&  \frac{\lambda - 2}{\lambda - 1} \cdot\frac{\lambda^\prime}{\lambda} \propto
1 + 24\sum_{n=0}^{\infty}\frac{n q^{2n}}{1 + q^{2n}} \;.
\end{eqnarray*}
$\beta_T$ has a Fourier expansion in $q^2 = \exp(2\pi i \sigma)$ because $T\in\Gamma_T$, while 
$\beta_R$ and $\beta_S$ are Fourier expanded in $q$ because the smallest translation in 
$\Gamma_R$ and $\Gamma_S$ is $T^2$.

When $a = 0,1,\infty$ the $\beta$-function degenerates to one of the Jacobi  $\theta$-functions,
\begin{eqnarray*}
\beta_{a=0} &\propto& \frac{\lambda^\prime}{\lambda(1 - \lambda)} =  i\pi\theta_3^4\\
\beta_{a=1} &\propto& \frac{\lambda^\prime}{\lambda}  =  i\pi \theta_4^4 \\
\beta_{a\rightarrow\infty} &\rightarrow& i\pi\theta_2^4 \; .
\end{eqnarray*}
Since these are finite on the interior of the fundamental 
domain the would-be quantum critical point has been pushed off the physical (finite)
part of parameter space. The uniqueness of the interpolating potential $\Phi_a$ in eq.\,(\ref{eq:Phia}) 
follows from the physical requirements we place on the analytic structure of the $\beta$-function. 
We want a simple zero at the delocalization critical point $\sigma_{\otimes}$ 
at a finite value of $\Im \sigma$, and the mildest possible singularities allowed by modular symmetry. 
Rankin's theorems \cite{Rankin} tell us that the weight of $\beta_\sigma$ can be carried by the logarithmic derivative
$\partial_{\sigma} \ln\lambda = \lambda^\prime/\lambda = i\pi \theta^4_4$ of the level two invariant $\lambda$, 
and that the remaining weightless factor is a rational function of that invariant.  By inspection of $\beta_R$, 
$\beta_S$ and $\beta_T$ it is clear that a fractional linear form is needed, but no more in order to avoid non-simple 
critical points.  In other words, 
$\beta_\sigma\propto F(\lambda)\partial_\sigma \ln\lambda  \propto \theta_4^4 F(\lambda)$ with 
\begin{equation*}
F(\lambda) = \frac{A\lambda + B}{C\lambda + D} \; .
\end{equation*}
Since the weight 2 form $\theta_4^4$ is finite (no zeros or poles) on $\mathbb H$,
any non-trivial analytic structure is carried by $F$.
The modular invariant $\lambda$ has a simple pole in $q$ on the fundamental domain $\mathbb F_2$. This cannot be avoided since the only holomorphic invariant is a constant. 
We now use the sum rule eq.\,(\ref{eq:sumrule}) for modular forms that 
transform with weight $w = 0$ or $w = 2$ under $\Gamma(2)$ ($\mu = 6$):
\begin{equation*}
n_0 + n_1 + n_{\infty}  + \sum_p n_p = \frac{w}{2} \; .
\end{equation*}
The possible zeros and poles of $F(\lambda)$ are at $\lambda = -B/A$ and $\lambda=-D/C$.
If the ``particle-hole" symmetry $\overline{\beta(\lambda)} =\beta(\overline{\lambda})$ holds, 
then $A,B,C,D$ are restricted to be real, and zeros and poles can only appear along the semi-circles where $\lambda$ is real \cite{BDD}. The requirement of having no poles except at the boundary $\overline{\mathbb H}_\oplus$ sets $-D/C =0$, $1$ or $\infty$. Similarly, the requirement of having a zero on $\mathbb{H} $, sets $-B/A\neq 0,1,\infty$.
If $D/C = -1$ there is a pole in $F$ at $\sigma = 0$, so $n_0(F) = -1$.  However, this does not affect the analytic structure since 
 $\theta_4^4$ has a zero at $\sigma = 0$, so $n_0(\theta_4^4) = 1$.  The pole from $F$ is therefore cancelled,
 $n_0(\beta_\sigma) = 0$, and this type of $\beta$-function can have the required critical point.
 
On the other hand,  if $D/C = 0$ there is a pole in $F$ at $\sigma = i\infty$, giving $n_{\infty}(F) = -1$, 
but since $n_{\infty}(\theta_4^4) = 0$ this contribution is not cancelled, $n_{\infty}(\beta_\sigma) = -1$, 
and there must be additional zeros to saturate the sum  rule for the $\beta$-function.  Since we only want one simple zero,
this is not a viable form of the $\beta$-function. 
Similarly, if $D/C = -\infty$ then $n_1(F) = -1$, and since $n_1(\theta_4^4) = 0$, this also requires additional zeroes and is therefore not a viable form for a physical $\beta$-function.

Since the metallic conductivity point ($\sigma\rightarrow i \infty$) should be a repulsive fixed point, 
the $\beta$-function has the form:
\begin{equation*}
\beta \propto -\frac{\lambda'}{\lambda(1-\lambda)}(A \lambda/B + 1) = -i\pi \theta_{3}^{4} (A\lambda/B + 1)\; , 
\end{equation*}
up to an undetermined positive multiplicative constant, which allows us to rescale $A/B = -a$, giving eq.\,(\ref{eq:betaa}).
This fixes $\beta_{\sigma}$ to be the gradient of the ansatz potential exhibited in eq.\,(\ref{eq:Phia}). 
 
When $a\neq\{a_R, a_T, a_S\} = \{-1, 1/2, 2\}$ the flow is only $\Gamma(2)$ symmetric, and the critical point with finite imaginary part ($\Im \sigma>0$) is not a fixed point of the group.   However, because $\lambda$ is real along the boundary of $\mathbb F_2$ \cite[p.228]{Rankin}, 
as long as $a\in\mathbb R$ the critical point is still on the semi-circular 
boundary of $\mathbb F_2$.  The semi-circle law observed in the spin-polarized case, 
for which $a = a_T = 1/2 \in\mathbb R$, is therefore an automatic consequence of modular symmetry, 
as pointed out in Ref.\,\onlinecite{BDD}.

\begin{figure}[h!]
\centering
\includegraphics[width=0.45\textwidth]{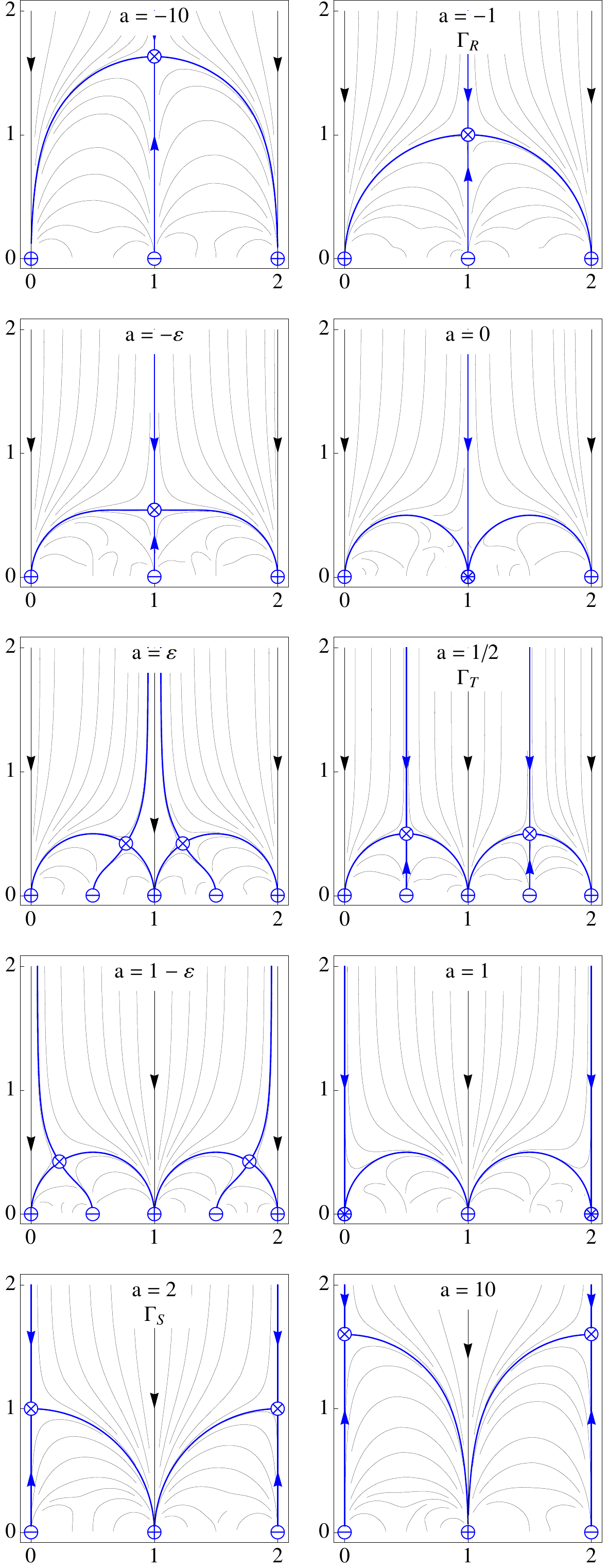}
\caption[multiplot]{(Color online)  $\Gamma(2)$-symmetric flows generated by $\Phi_a$ 
for ten values of $a\in [-10,10]$ ($\varepsilon = 0.05$).   At $a = -1, \,1/2$ and $2$ 
the symmetry is enhanced to $\Gamma_R$, $\Gamma_T$ and $\Gamma_S$. Thick blue lines are phase boundaries or separatrices.}
\label{fig:Fig5}
\end{figure}

Fig.\,\ref{fig:Fig5} shows a sequence of ten distinct RG flows, obtained for real values of 
the parameter in the range $a \in[ -10, +10]$, with the choice $i\infty = \ominus$
(if $i\infty = \oplus$ the direction of all arrows is reversed).
The location of the quantum critical point $\otimes$ (and some of its modular reflections) 
is computed using eq.\,(\ref{eq:qcp}), and as expected they move along boundaries of fundamental domains for $\Gamma(2)$.  Thick (blue) lines are phase boundaries or separatrices.

For increasing values of $a < 0$ the quantum critical point $\otimes$ first
moves straight down the vertical line $\Re \sigma = 1$
($a =  -10,\, -1,\, -\varepsilon = -0.05$), until it fuses with the repulsive point $\ominus = 1$ 
as $a\rightarrow 0^-$, denoted by $\ostar = 1$ in the $a = 0$ diagram.  
A further increase in $a$ appears to split $\ostar$ into three parts ($a = \varepsilon$):
a mirror pair of critical points $\otimes$ that move apart along semi-circles
(as described in detail in Sect.\,2.1: $0^+\leftarrow \otimes$, $T^2R^2(\otimes)\rightarrow 2^-$), leaving behind a new attractive point $\oplus = 1$. 
This attractor is attached to a new phase injected during the split, 
which squeezes out the old phases 
($a = 1 - \varepsilon$) as $a$ increases and eventually 
engulfs much of the diagram ($a = 1$).   At $a = 1/2$ this new phase provides 
the ``real estate" needed to enhance the translation symmetry from $T^2$ to $T$ 
(and the total symmetry from $\Gamma(2)$ to $\Gamma_T$).
When the parameter reaches the value $a = 1$ the exact opposite of a split occurs: 
mirror pairs of critical points coalesce at $\ostar = 0$, and then move straight up 
($a = 2,\,10$), leaving behind a repulsive fixed point $\ominus = 0$.

Notice how this resolves the enigma of having a discontinuous ``jump" in symmetry
at $a = -1,\,1/2$ and $2$, which at first sight seems mysterious and unphysical, 
especially since the model acquires an infinite number of new non-commuting symmetries.  However, our discussion of the RG potential $\Phi_a$, which generated the flows shown in 
Fig.\,\ref{fig:Fig5}, makes it transparent that this ``spontaneous symmetry generation" 
is no more mysterious, discontinuous or ``spontaneous" than ordinary 
``spontaneous symmetry breaking". 
The potential changes continuously and smoothly as a function of $a$, interpolating
between different (infinite and non-abelian) symmetries by switching parts of 
the potential on or off, thus injecting, expanding or shrinking phases as needed \cite{L1}.

Some of the diagrams in Fig.\,\ref{fig:Fig5} are similar to those appearing in 
Ref.\,\onlinecite{D1}, where a different interpolation between enhanced symmetries based on 
``scaling functions''  is investigated.  We defer a more detailed comparison to 
 Sect.\,\ref{sec:otherwork}.

\section{Comparison with experiment}
\label{sec:comparison}
\noindent
We wish to confront the proposed RG potential $\Phi_a$ (eq.\,(\ref{eq:Phia})) 
with available scaling data.
The simplest cases, when the spin is fully polarized or fully resolved, have previously been
identified as having the enhanced symmetries  $\Gamma_T$  $(a_T = 1/2)$  and
$\Gamma_R$ $(a_R = -1)$ \cite{LR1,L1,LR2,LR4}. 
The RG flows generated by $\Phi_{1/2}$ and $\Phi_{-1}$ were found to be in excellent 
agreement with available scaling data for both integer and 
fractional transitions \cite{Grenoble1, Grenoble2} .

Here we extend this analysis to more general situations that
allow us to test if the proposed potential does interpolate between 
the points of enhanced symmetry considered previously. 
Since we shall be discussing both flows in the complex conductivity parameter $\sigma$
and resistivity parameter $\rho = S(\sigma)$, when confusion can arise the critical point 
$\otimes$ and the parameter $a$ will be indexed accordingly.  

We are not at this time able to derive the value of $a$ from micro-physical considerations.
The question we address here is whether it is possible to parametrize all, or almost all, 
scaling data with a single real number $a$ that fixes the modular RG potential $\Phi_a$\,, 
and therefore the location of all quantum critical points and the geometry of all RG flow lines.
This ``phenomenological" approach, explained at length in the introduction, is extremely rigid
and therefore eminently falsifiable.  

\begin{figure}[t]
\centering
\includegraphics[width=0.46\textwidth]{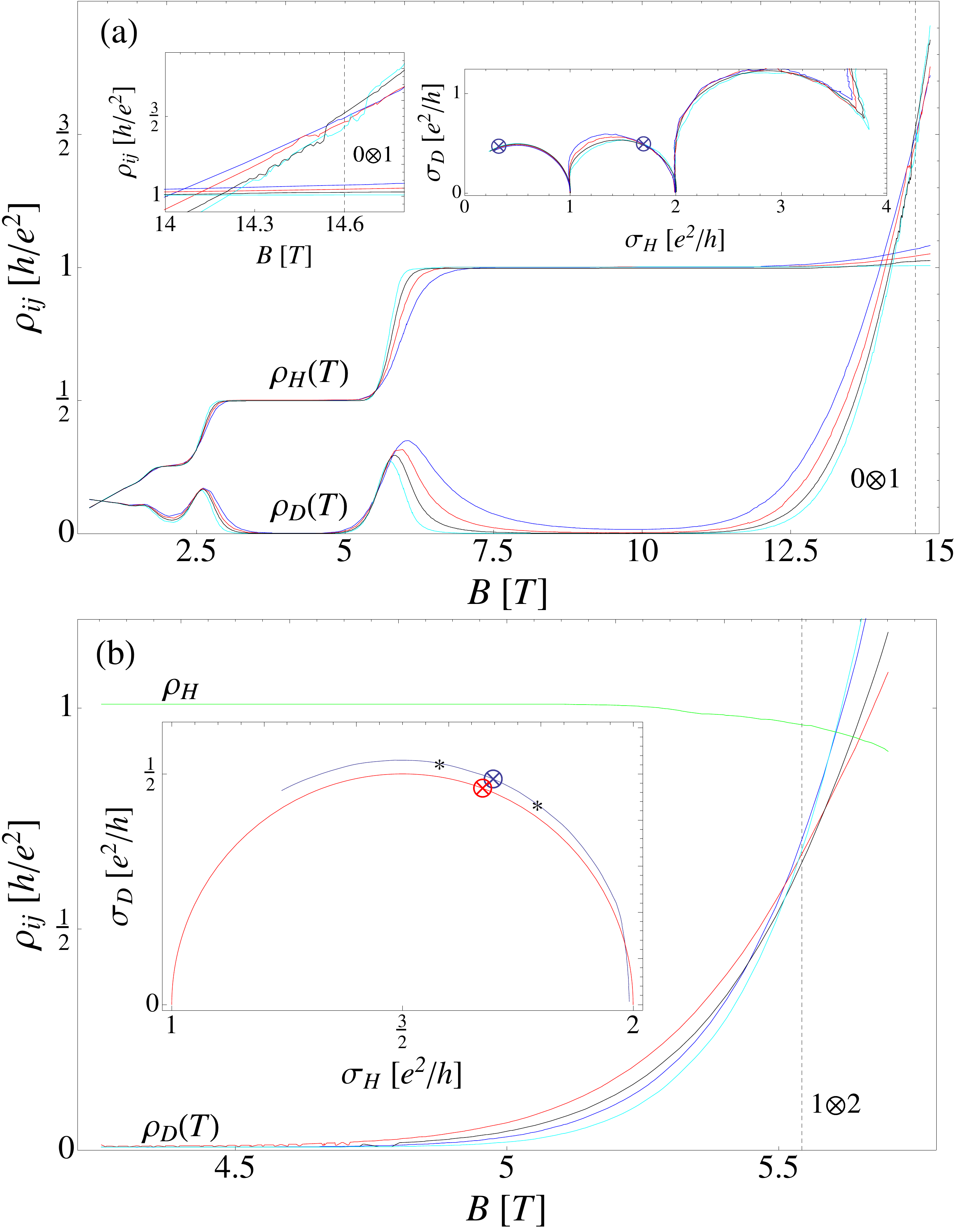}
\caption[doubleplotgivingfontpoblem]{(Color online) Temperature driven flow data 
\cite{Huang2004} for
(a) the $0\otimes_\sigma 1$ transition, and (b)  the $1\otimes_\sigma 2$ transition, 
in an AlGaAs/GaAs heterostructure with density 
$n= 2\times 10^{11} /cm^2$ and mobility $\mu = 7.6 \times 10^3 cm^2/Vs$.
The insets show the location of the quantum critical points in the conductivity 
phase diagram, compared with the critical points of  
the interpolating RG potential $\Phi_{0.89}$.}
\label{fig:Fig6}
\end{figure}

\subsection{Spin split Landau level}\label{sec:resolved}
\noindent
Huang et al.\,\cite{Huang2004} have studied the conductivity transition between the first and second Hall level, which we denote by $1\otimes_\sigma 2$,  and also the $0\otimes_\sigma 1$
transition, in a semiconductor sample where the spin degeneracy 
of Landau levels is small but resolved. The gap between the two spin flavors is 
determined by the ratio of the electron effective mass and effective $g^{*}$-factor 
in the sample. The sample is a AlGaAs/GaAs heterostructure with density 
$n= 2\times 10^{11} /cm^2$ and mobility $\mu = 7.6 \times 10^3 cm^2/Vs$. 

For the  $0\otimes_\rho 1$ transition they reported a 
 a critical magnetic field value $B_\times \approx 14.7\,T$, a critical value of the diagonal 
 resistivity $\rho_D(0\otimes 1) \approx 37\,k\Omega \approx 1.43\,[h/e^2]$, and 
 that $\rho_H$ appears to be independent of $B$ for the lowest temperatures
 (see  Fig.\,\ref{fig:Fig6}(a)).
 
However, the crossing point appears to be moving down with decreasing temperature,  
and we estimate that $B_\times (0\otimes 1)\approx 14.6\pm 0.1\,T$, which gives 
the critical value $\rho_\otimes \approx 1.007 + 1.439 i \,[h/e^2]$ for the 
$0\otimes_\rho 1$ transition.
 This is very close to the value $\rho_\otimes \approx 1 + 1.439 i \,[h/e^2]$ 
 obtained from $\Phi_a$ with $a_\rho \approx -5.25$. 
 The corresponding value for the conductivity flow parameter is 
 $a_\sigma \approx 0.84$, and the phase diagram is shown as an inset in 
 Fig.\,\ref{fig:Fig6}(a).
 
For the  $1\otimes_\rho 2$ transition they find that the critical magnetic field value is
$B_\times \approx 5.55\,T$.   This gives a critical point on the semi-circle 
shown in the inset in Fig.\,\ref{fig:Fig6}(b), which is almost identical to the critical 
point of $\Phi_{0.84}$ also shown in the same diagram.

\begin{figure}[t]
\includegraphics[width=0.475\textwidth]{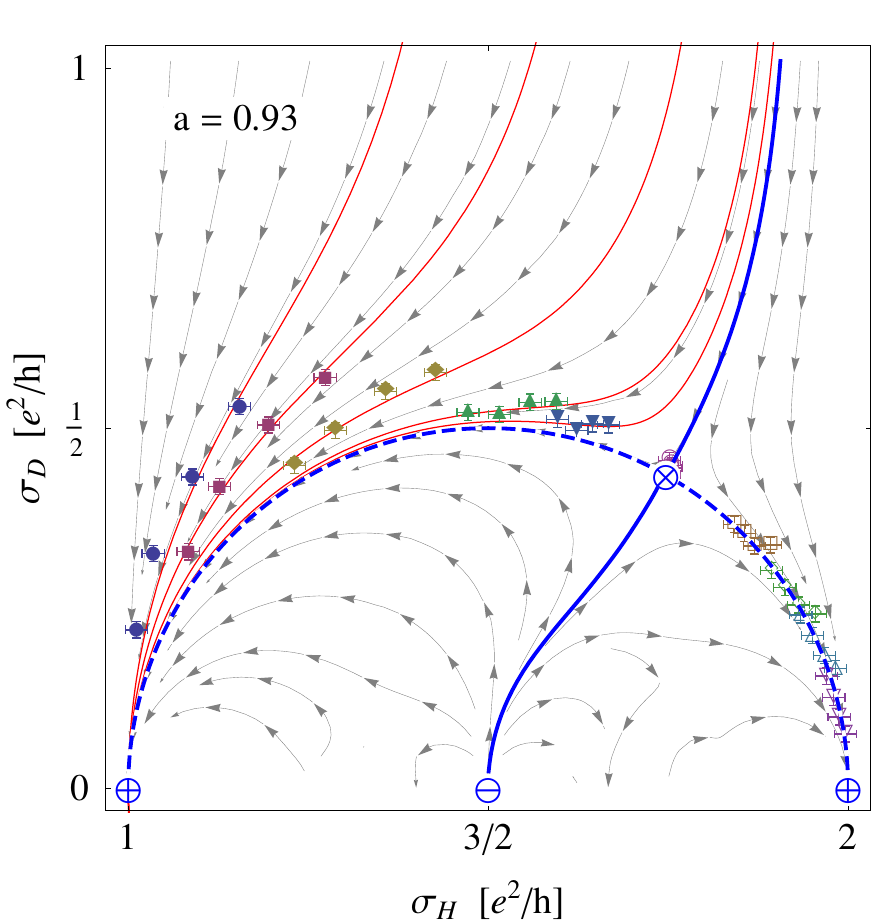}
\caption[excellentfit]{(Color online) Temperature driven flow data for the $1\otimes 2$ 
conductivity transition (adapted from Ref.\,\onlinecite{Huang2007}), compared with the 
theoretical flow generated by the interpolating RG potential $\Phi_{0.93}$. 
Thick (blue) lines are theoretical phase boundaries. Dashed blue lines are separatrices.}
\label{fig:Fig7}
\end{figure}

\begin{figure}[t]
\includegraphics[width=0.475\textwidth]{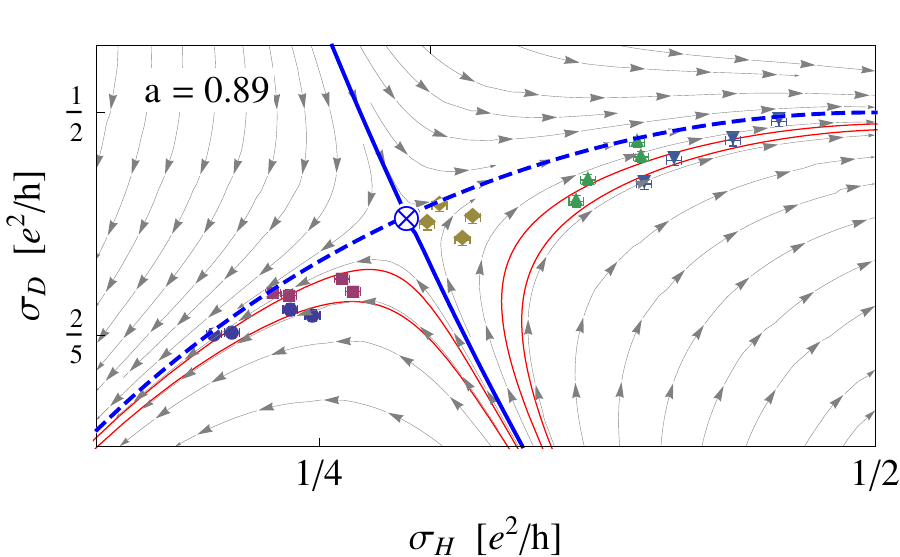}
\caption[tolerablefit]{(Color online) Temperature driven flow data for the $0\otimes 1$ 
conductivity transition (adapted from Ref.\,\onlinecite{Huang2007}), compared with the 
theoretical flow generated by the interpolating RG potential $\Phi_{0.89}$.  
Thick (blue) lines are theoretical phase boundaries. Dashed (blue) lines are separatrices.}
\label{fig:Fig8}
\end{figure}

These results appear to show that there is a $\Gamma(2)$ symmetry emerging in this sample,
with an RG potential $\Phi_{0.84}$ whose critical points are close to the observed values
for the $0\otimes_\sigma 1\otimes_\sigma 2$ transitions.
Using the same sample as in the experiment described above \cite{Huang2004}, 
Huang et al.\,\cite{Huang2007} have studied the temperature scaling in more detail.
The experiment probed temperature driven RG flows, with temperatures sampled at
$T = 0.94, 0.68, 0.49, 0.31\, [K]$, and their data are reproduced in Figs.\,\ref{fig:Fig7} 
and \ref{fig:Fig8} (adapted from Ref.\,\onlinecite{Huang2007}).

The semi-circular geometry of the flows strongly suggests that the flow is following the separatrix connecting the plateaux at $\sigma_\oplus = 1$ and $\sigma_\oplus = 2$, if we
 allow for a systematic error that shifts $\sigma_D$ up by $0.17\,e^2/h$, in which case 
 a real value of the parameter $a$ will suffice.  This value is most easily determined by finding the location of the quantum critical point $\sigma_\otimes$, and comparing this with the best fit to 
 $\Phi_a$.  If $\sigma_\otimes$ were known exactly we could simply compute the value of $a$
using eq.\,(\ref{eq:qcp}), i.e., $a = 1/\lambda(\sigma_\otimes)$.  However, with the substantial uncertainties in currently available data it is often better to make a global comparison
of all the scaling data, not just those near the critical point, in order to find a reasonable 
value for $a$ giving a good overall fit.  

For the $1\otimes_\sigma 2$ transition we find that  $a = 0.93\pm0.01$ 
gives a good fit to the data, as shown in Fig.\,\ref{fig:Fig7} where the flow generated by  
$\Phi_{0.93}$ has been superimposed on the data from Ref.\,\onlinecite{Huang2007}.
Notice that there is a cluster of data that is almost stationary, 
and therefore should be very close to an RG fixed point.
This is indeed the case, as the global data fit gives a quantum critical point $\otimes$ 
that almost totally eclipses this data series. 

For the $0\otimes_\sigma 1$ transition we find that  $a  = 0.89\pm0.04$ 
gives a reasonable fit, as shown in Fig.\,\ref{fig:Fig8} where the flow generated by  
$\Phi_{0.89}$ has been superimposed on the data from Ref.\,\onlinecite{Huang2007}.
Notice that there is also in this case a cluster of data that is more or less stationary, 
albeit not as convincingly as in the previous diagram.  
These data points are indeed quite close to the candidate critical point $\otimes$ 
identified from the rest of the flow data, as expected if the global fit is reasonable.

In both diagrams the thin solid (red) flow lines have been chosen to illustrate that the data are 
 consistent with the potential.  In Fig.\,\ref{fig:Fig7} the fit is essentially perfect,
even with the small error bars we have estimated (since none were published).
Comparing these results with Fig.\,\ref{fig:Fig6},
there appears to be a $\Gamma(2)$ symmetry emerging in this sample,
with an RG potential $\Phi_a$ consistent across several quantum Hall transitions.
The small dispersion in $a$-values is not surprising given the precision of these experiments.
Since we estimate that $B_\times(1\otimes2) \approx 5.54 \pm .05\,T$, the uncertainty in 
 $a_\sigma$ is $\pm 0.1$ in both transitions.

In addition to the the flows presented here, the transition $2\otimes_\rho 4$
was also investigated in Ref.\,\onlinecite{Huang2007}.  They found a critical point 
at $\rho_\otimes = (7.75, 3.54) \pm 0.05 \,k\Omega \approx (0.300,0.137)\pm0.002\, [h/e^2]$,
which corresponds to $a_\rho = -1.76$. The authors remark that the plateau in 
$\rho_D$ is not well developed, and note that if we assume that the true value is 
$\rho_D = 1/10\,[h/e^2]$, then $\rho_H = 3/10\,[h/e^2]$ (so $\rho_\otimes = (3 + i)/10\,[h/e^2]$) corresponds exactly to the parameter value $a_\rho = -1$, 
i.e., the enhanced symmetry $\Gamma_R$ for the 
usual spin unresolved quantum Hall effect, 
which is often observed at higher Landau levels.

\subsection{Zeeman splitting}
\begin{figure}[h]
\centering
\includegraphics[width=0.48\textwidth]{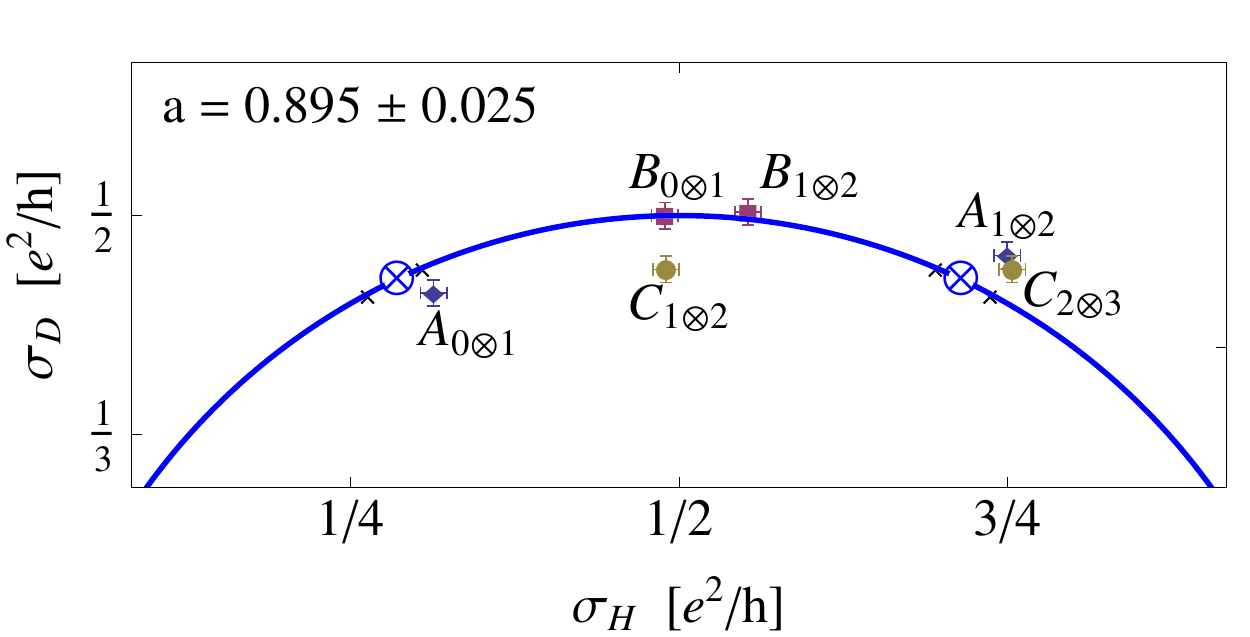}
\caption{(Color online) Critical points for the transitions studied in the samples $A$, $B$ and $C$
discussed in the text, are in rough agreement with the generalized semi-circle law
(data adapted from Ref.\,\onlinecite{Hang2005} Fig.\,3).  In order to make a global comparison
some of the data have been rescaled and translated to the proximity of the exhibited semi-circle. 
The margin of error is indicated by black crosses.}
\label{fig:Fig9}
\end{figure}

\noindent
Hang et al.\,\cite{Hang2005} have studied the effect of Zeeman splitting in IQHE transitions in three different samples (called $A$, $B$ and $C$) with various spin degeneracies in the Landau spectrum. Sample $A$ is the same sample as in Refs.\,\onlinecite{Huang2004, Huang2007}, analyzed in the previous section. Sample $B$ is a GaAs system with InAs dots, but neither the  mobility nor the density have been reported.  Sample $C$ is a SiGe hole system with mobility $\mu = 1.69 \times 10^4 cm^2/Vs$ and density
$n\approx 2.2 \times 10^{11}/cm^2$. Their data are reproduced in Fig.\,\ref{fig:Fig9}.
While the data are less than ideal, they do appear to provide some evidence for a 
\emph{generalized semi-circle law},  which asserts that all quantum critical points should appear on the boundary of $\mathbb F$.  This is an automatic property of $\Phi_a$ flows if $a$ is real.

Only the plateau at $\rho_\oplus \approx 2$ is properly developed, while the second visible plateau 
at  $\rho^H_\oplus \approx 1$ does not have a fully developed zero of $\rho_D$, indicating that the 
scaling regime has not been reached.
 Accordingly, the transitions deviate somewhat from the expected semi-circles, although they do appear to indicate  the common trend, with two transitions above and two under the semi-circle. 
 The critical points are on the semi-circles to a reasonable accuracy, within a couple of the rather 
 large standard deviations in these experiments.  

Their conductivity data allows us to compare the location of the critical points in the transitions 
$0\otimes_\rho 1\otimes_\rho 2$ for sample $A$, giving
$0\otimes_\rho 1\approx (0.31, 0.44)\pm 0.01$ and $1\otimes_\rho 2 \approx (0.75, 0.47) \pm 0.01$. These correspond fairly well to the value $a_{\sigma}= 0.895\pm 0.025$, in agreement with the estimates
$a_{\sigma}\approx0.93\pm0.01$ and $a_{\sigma}=0.84\pm0.1$ obtained for this sample in the previous section. 

The critical points for the other samples are also approximately on the semi-circle. 
For sample $C$ our estimates for the value of $a$ 
are different for the $1\otimes2$ and $2\otimes3$ transitions, but this is to be expected since the spin configuration drastically lowers the the splitting between higher Landau levels 
(see Ref.\,\onlinecite{Hang2005}, Fig.\,1).  The $2\otimes 3$ transition therefore only enjoys a $\Gamma(2)$ symmetry, whereas the the well resolved transition $1\otimes 2$ obeys the full $\Gamma_T$ symmetry. 

For sample $B$ similar remarks apply, but in this case the $0\otimes1$ transition reflects the spin-split symmetry $\Gamma_T$, while the $1\otimes   2$ transition reflects the spin-unresolved symmetry $\Gamma_R$ (compare Ref.\,\onlinecite{Hang2005}, Fig.\,1). 
 As the best fit to the data we find $a_{\sigma}\approx 0.52\pm 0.4$ for the $0\otimes 1$ transition.
From the $1\otimes 2$ data we extract $a_{\sigma} \approx -0.94 \approx -0.94 + 0.46 i$.
This means that the flow is very close to the $a_{\sigma} = -1$ flow near the semi-circle, 
and that the complex part of $a_{\sigma}$ only displaces the critical point and the flow 
slightly from the centre of the semi-circle.

We interpret these experiments as providing supporting evidence for the generalized semi-circle law.
For sample $A$ sufficient data are available for us to extract $a_{\sigma} = 0.895\pm 0.025$ 
from the location of the critical points, and this is consistent with  $a_{\sigma}\approx 0.93\pm0.01$ and $a_{\sigma}=0.84\pm0.1$, obtained in the previous section from the data in 
Refs.\,\onlinecite{Huang2004, Huang2007}.

\subsection{Overlapping spin sub-bands}

\begin{figure}[t]
\includegraphics[width=.485\textwidth]{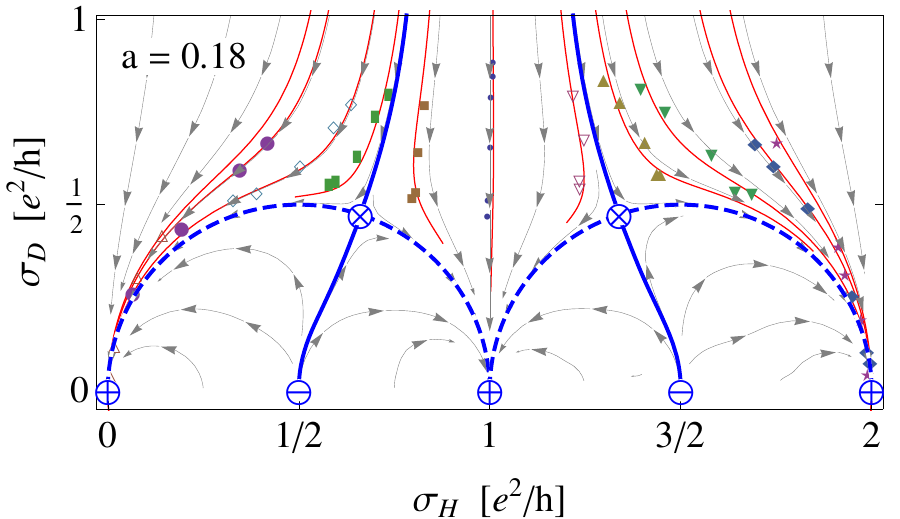}
\caption{(Color online) Low temperature flow data for the 
$0\otimes_\sigma1\otimes_\sigma2$ transitions in sample S34 
(adapted from Ref.\,\onlinecite{Grenoble3} Fig.\,2), 
compared with the theoretical flow generated by  
the interpolating RG potential $\Phi_{0.18}$.  
Thick (blue) lines are theoretical phase boundaries. Dashed (blue) lines are separatrices.}
\label{fig:Fig10}
\end{figure}

\noindent
Murzin et al.\,\cite{Grenoble3} have studied a quantum Hall system in which the spin splitting is 
small enough for spin split Landau levels to overlap substantially. Their three samples 
(labelled S34, S40 and S50) are heavily Si-doped GaAs layers with mobilities 
$\mu\approx (2.0 - 2.4)\times 10^3\, cm^2/Vs$ and densities $n = 3.9, 4.6, 5.0 \times 10^{11}/cm^2$. 
They estimate the Zeeman splitting to be of the order of $4\,K$, 
whereas the level broadening is much larger, of the order of $100\,K$.  

Data obtained above a few hundred milli-Kelvin are not expected to be in the scaling domain,
so we disregard the high temperature data (which start at $T \approx 12\,K$).
The situation is quite similar to the original experiment where scaling was first observed in 
the QHE \cite{Tsui2}, and where this procedure was justified, 
but it has only recently been pointed out that scaling data exhibit a 
hyper-sensitivity to temperature that still does not receive sufficient attention \cite{LR2}.

The low-temperature data (down to $T = 0.1\,K$) are reproduced in Fig.\,\ref{fig:Fig10} 
(adapted from Ref.\,\onlinecite{Grenoble3}), and compared to the flow generated 
by $\Phi_a$ with $a = 0.18\pm 0.01$.  Since the experimental errors (which are not reported) must be
at least as large as the plot markers used in Ref.\,\onlinecite{Grenoble3}, Fig.\,2, 
there is no discrepancy between the data and the theoretical flow.

The theoretical critical points $0\otimes_\sigma1 \approx 0.66 + 0.47i$ and 
$1\otimes_\sigma2 \approx 1.34 + 0.47i$ are reasonably close to the corresponding 
experimental values $\sigma_{0\otimes1} = (0.5\pm 0.05, 0.5\pm 0.02)\,[e^2/h]$ and
$\sigma_{1\otimes2} = (1.5\pm 0.05, 0.5\pm 0.02)\,[e^2/h]$, based on a linear extrapolation 
of the $\sigma_D$-maxima to zero temperature \cite{Grenoble3}.

In Ref.\,\onlinecite{Grenoble3} it was suggested that their data could be interpreted as a 
hybrid flow composed of two spin components scaling independently. 
However, the good agreement we have found with the $\Gamma(2)$ symmetric potential 
suggests that the observed scaling simply follows from reduced spin splitting that leads 
to near degenerate Landau levels.  The most obvious consequence of this re-interpretation is 
that the location of the quantum critical points is shifted away from the more symmetric spin-polarized 
positions, as shown in Fig.\,\ref{fig:Fig10}.
This should be straight forward to verify by more accurate experiments.

\subsection{Layered heterostructure}
\noindent
Huang et al.\,\cite{Huang2003} have studied the $0\otimes_\rho1\otimes_\rho2$ transitions in a layered AlGaAs/InGaAs/GaAs heterostructure with mobility $\mu = 1.0\times 10^5 cm^2/Vs$ and carrier density
$n = 1.3 \times 10^{11}/cm^2$.  Again the high-temperature data (above $1\,K$) are not relevant 
for our study of the quantum phase transitions.

For the $1\otimes_\rho2$ transition they observed a stationary flow that they interpreted as being 
controlled by a critical point  $\sigma_\otimes \approx 1.9 + 0.39i \,[e^2/h]$. 
In comparing this result with the RG potential $\Phi_a$, the best value we have found is 
$a_\sigma \approx 0.992$, which gives $\sigma_\otimes \approx1.85 + 0.35 i$. 
However, in this case the experimental flow does not fit the theoretical flow.
To avoid having experimental and theoretical flow-lines crossing, 
the data might be conjectured to fall on a semi-circular separatrix, 
but then the ``semi-circle" formed by the experimental data points would be 
skewed and distorted, so this seems implausible. 

For the $0\otimes_\rho1$ transition they report 
$\rho^D_\otimes \approx 20\,k\Omega \approx 0.77\,[h/e^2]$, and
$0 \otimes_\rho 1\approx 1 + 0.77 i\,[h/e^2]$.  This gives 
$0\otimes_\sigma 1\approx 0.62 + 0.48i$, which corrresponds to $a_\sigma \approx -0.32$, 
in contradiction with the value $a\approx 0.992$ we obtained above for the $1\otimes 2$ transition.  The critical points of the two transitions appear at values of $\sigma_H$ that 
are too high for them to be related by $\Gamma(2)$ symmetry, i.e, they are not 
symmetric with respect to the $\sigma_H = 1$ line.

The authors of Ref.\,\onlinecite{Huang2003} comment that these rather unexpected results could be due to a variety of reasons: inhomogeneity in the sample, mixing between Landau bands, small spin-splitting, or temperatures too high for scaling to be observed.   
In any case, this system does not appear to belong to the family of  RG flows generated by 
$\Phi_a$, and therefore highlights the need for a better understanding of necessary and sufficient 
conditions for the emergence of modular symmetries.

\subsection{Remarks on graphene}
\noindent
A single layer ($b = 1$) of graphene has a quasi-particle spectrum with a ``relativistic" dispersion relation, i.e., they are essentially Dirac fermions. Bilayer ($b = 2$) graphene has the spectum of massive Dirac fermions. The free particle Landau problem gives integer quantum Hall (IQH) plateaux at:
\begin{equation}
\sigma_\oplus = \sigma_H = 4 (N + \frac{b}{2}) \; ,\quad N\in \mathbb Z,
\label{eq:grapheneIQH}
\end{equation}
in natural units ($e^2/h$).
The factor 4 arises from two-fold spin and valley degeneracies in 
graphene \cite{GraphIQH}. 

As originally noted in Ref.\,\onlinecite{BD1}, if $\sigma$ is replaced by $\sigma/2$
in order to take account of either the  spin- or valley degeneracy, then the fixed point structure of the QHE in single-layer graphene 
appears to respects the modular symmetry $\Gamma_S$. 
The system can be argued to be a spin-polarized system, but
because of a  Berry phase $\pi$ picked up by Dirac fermions moving around the Dirac cone \cite{GraphIQH, BD1}, the group $\Gamma_T$ should be 
conjugated by a $\pi$-flux attachment $STS$, which gives the bosonic duality group 
$\Gamma_S$ \cite{LR1,L1, BD2}.  
This reduces the translation symmetry from $T$ to $T^2$, and this leads to the correct 
periodicity of $4$ in eq.\,(\ref{eq:grapheneIQH}).

In bilayer graphene the Berry phase is $2\pi$, which gives the symmetry $\Gamma_R$ 
after the conductance is divided by 2. This is consistent with eq.\,(\ref{eq:grapheneIQH}), 
except that the $\sigma_\oplus = 0$ state is missing.  Instead there is a metallic state with a minimum conductivity.  The zeroth Landau level state can be created by doping.

Recent advances in graphene sample fabrication have revealed fractional quantum Hall (FQH) states \cite{GraphFQH}, as well as anomalous IQH states that do not fit 
into the non-interacting sequence given by eq.\,(\ref{eq:grapheneIQH}). 

The anomalous states have been related to symmetry breaking between the spin- and valley degeneracies of Dirac fermions at high magnetic fields \cite{GraphBreak}, and
by now almost all IQH states have been observed in graphene devices at high magnetic fields
\cite{GraphFQH, GraphAnomaly, Graph8, GraphBi}. 
In particular, all the broken symmetry states of the $N=0$ Landau level, 
giving plateaux at $\sigma_\oplus = 0, 1, 2, 3$ and $4$, have now been observed for both 
single- and bilayer graphene \cite{Graph8,GraphMulti}.
There is also evidence for all plateaux $\sigma_\oplus = 1,\dots, 16$ for bilayer graphene \cite{GraphBi}, which suggests that the strong field has completely ``polarized" the system, lifting the multicomponent degeneracy and returning the symmetry to the conventional one-component group $\Gamma_T$.
Recently, Ref.\,\onlinecite{GraphMulti} reported fractions that are predominantly multiples 
of $1/3$, although some states are missing from the $N=0$ Landau level.

Unfortunately scaling data for the plateaux transitions in graphene are not yet available. 
There is some evidence that the plateau-plateau transitions are in the same universality class 
as in the usual spin-polarized QHE, reported in Refs.\,\onlinecite{GraphPlat,GraphIns},
but they also find evidence that the insulator-plateau transition is in a different universality class.

Nevertheless, the fixed points of the flows appear to be consistent with the symmetry
transmutation \cite{Rankin, L1} $\Gamma_S \xrightarrow{\sigma/2 \to \sigma} \Gamma_T$
driven by an increasing magnetic field, leading eventually to a single flavor of spin-polarized fermions without the $\pi$-Berry phase due to a (spontaneous) mass gap. 
Clearly this kind of a scenario (or something more complicated) can be achieved with the proposed ansatz, where $a$ is a non-universal sample specific parameter that depends on
impurities, sample chemistry, sample architecture, strength of the external magnetic field,
 Zeeman splitting, valley separation, etc.

Finally we remark that if low energy models of graphene do enjoy modular symmetries,
then they can in principle be used as a diagnostic tool to characterize the material.
 Abanin and Levitov \cite{AbaninLevitov} have recently investigated another
  way to use transport measurements 
to characterize materials exhibiting the quantum Hall effect, including the 
number of layers in a graphene sample.  Since the total macroscopic 
conductance measured in the two-terminal experiments considered by 
these authors depends on sample geometry (e.g., the aspect ratio), it is 
conceivable that this will conceal the properties of the material (e.g., the 
electron density) and therefore render the method unsuitable for sample 
characterization. Using a semi-circle relation between the conductivities they 
show that this is not the case, so that their method can be used to 
determine the number of layers in graphene, for example.

\section{Comparison with other work}
\label{sec:otherwork}
\noindent
There have been various attempts to analyze  the phase structure of the QHE,
starting with a proposal by Khmel'nitzkii \cite{Khmelnitzkii} that identified
a translation symmetry of RG flows in the IQHE. This was
motivated by a sigma-model of localization, which was
extensively elaborated by Pruisken et al.\,\cite{Pruisken1,Pruisken2}, but the target space
geometry does not appear to be rich enough to include the FQHE.

Ref.\,\onlinecite{LR1} proposed that a modular symmetry would be capable of describing both the integer and the fractional Hall effects by including dualities in addition to translations, as described in the introduction.  The three maximal index 3 subgroups  $\Gamma_X\, (X = R, S, T)$ 
of the full modular group ${\rm PSL}(2,\mathbb Z)$ were immediately identified as the largest symmetries of physical interest.  The group ${\rm Aut}\Gamma_T$,
acting by holomorphic fractional linear transformations on the complexified conductivity, was shown to give the correct phenomenology (fixed point structure and super-universality) 
of the spin polarized QHE.
Since $\rho = S(\sigma)$ this is equivalent to having an emergent 
 $\Gamma_R$ symmetry (which is $S$-conjugate to $\Gamma_T$) 
 on the space of resistivities, and $\Gamma_S$ was proposed as the relevant symmetry 
 for analogous transport problems with bosonic quasi-particles \cite{LR1}.
 
The symmetry $\Gamma(2)$ and interpolations between the
enhanced symmetries $\Gamma_X\,(X = R, S, T)$ were introduced in Ref.\,\onlinecite{L1},
where the idea of an interpolating potential elaborated here was also proposed.
 
At  roughly the same time superficially similar dualities \cite{KLZ, WZK, Jain} acting 
on the filling fraction $\nu$ were considered, and the resulting transformations are known collectively as ``the law of corresponding states".  Since $\nu$ is essentially the plateau value
$\sigma_\oplus = \sigma^\p_H \in \mathbb R$, this approach is oblivious to the complex 
structure that gives modular symmetries most of their predictive power.
These dualities appear to disagree with experiment, unlike the complexified duality 
identified in Ref.\,\onlinecite{LR1} that is in excellent agreement with available data \cite{LR2,LR4,LR5}.

Following the ideas in Ref.\,\onlinecite{KLZ}, Burgess et al.\,\cite{BD1} related complexified 
modular transformations to dualities in the low energy (linear response) 
transport coefficients of an effective Chern-Simons Landau-Ginzburg field theory.
These dualities follow from periodicity of the statistics angle of the charge carriers in 
two dimensions, together with particle vortex duality, 
assuming that only particles or only vortices take part in low energy charge transport. 
This gives $\Gamma_T$ symmetry for fermionic charge carriers, and 
$\Gamma_S$ symmetry for bosonic carriers. 
Note that dualities relate systems with different (anyonic) statistics of charge carriers,
but never bosonic charge carriers with fermionic ones. These dualities were further 
extended to the non-linear regime of charge transport in Ref.\,\onlinecite{BD3}. 

Burgess et al.\,\cite{BD4} have also considered generalized modular symmetries acting on bilayer systems, assuming that each single layer systems exhibits emergent modular symmetry in its scaling behavior. In the most general case a bilayer system has two independent complex conductivities, one for each layer, plus an inter-layer conductivity.  A genus two modular group $Sp(4, \mathbb{Z})$ then acts on a complex two-dimensional matrix of conductivities 
(with positive imaginary part).
If the bilayer system consists of identical layers, then the scaling behavior is
expected to be the same as in a mono-layer system,
and  the experimentally accessible total conductivity $\tilde\sigma$ 
for the bilayer is just the double of the conductivity  $\sigma$ for a single layer \cite{BD4}. 
If the scaling properties of a system with a single fermionic degree of freedom is determined by 
$\Gamma_T$, then $\tilde\sigma = 2\sigma$ implies that the scaling of a bilayer system is constrained by $\Gamma_R$. One example of this is observed when a spin polarized system changes to a spin degenerate system, summarized by the symmetry 
transmutation $\Gamma_R \xrightarrow{\sigma\to \sigma/2} \Gamma_T$ that is
induced by decreasing the magnetic field \cite{Rankin, L1}.
On the other hand, if the scaling of a single bosonic degree of freedom is determined by
$\Gamma_S$, then the scaling of the total conductivity $\tilde\sigma$ 
of a bilayer system should respect a symmetry generated by 
$T^4(\tilde\sigma) = \tilde\sigma +4$ and
$\tilde S(\tilde\sigma) = - 4/\tilde\sigma$.  This is verified by using the duality
 transformations $T^2$ and $S$ acting on $\sigma = \tilde\sigma/2$.   
Although $\tilde S$ is not a modular transformation,  the substitution  
$\tilde\sigma \rightarrow \sigma$ should produce a modular scaling diagram.
Ref.\,\onlinecite{BD4} compares these predictions with the fixed point structure observed in multi-component quantum Hall experiments.  While the experimental situations 
is still quite murky, there does not appear to be any inconsistency.

Dolan \cite{D1}  has argued that one should expect the 
duality group  to be $\Gamma(2)$ in systems with a small but 
non-vanishing difference between adjacent Landau levels of different flavors (spin, layer, etc.),
so that there is mixing between the Landau levels.
The degenerate bilayer case is the limit of exact degeneracy, and the single layer case the limit where there is no mixing between Landau levels. The transformation $ST^2S$ is the 
$2\pi$-shift (or flux attachment) in the statistics of the charge carriers \cite{BD1}, and should therefore always be
present in two dimensions. The translation $T^N$ counts the degeneracy of Landau levels, since the Hall conductivity transforms as $T^N(\sigma^\p_H) = \sigma^\p_H + N$. 
The symmetry is therefore $\Gamma(2)$ for fermionic charge carriers if one assumes that the chemical potential cannot be tuned in the gap between the different flavors of Landau levels. 
If both particles and vortices carry charge the situation is more complicated, but 
since $ST^2S$ is the periodicity in the statistics angle and one expects that there is no 
mixing between Landau levels of a single flavor of charge carriers, 
one should expect that the charge transport phenomena in two dimensions is at  least 
$\Gamma(2)$-symmetric.

There have been two other attempts to construct families of interpolating $\beta$-functions  
for the QHE  \cite{D1,D2,GMW}.
They have both retained the original idea from Ref.\,\onlinecite{LR1}, elaborated in 
Ref.\,\onlinecite{BL1}, that the physical (contravariant) $\beta$-function should be automorphic
under modular transformations with weight $w = -2$.  This is sufficient to give pictures of automorphic vector fields that look similar to ours, since any automorphic meromorphic 
function will by definition have this property.   They do not, however, consider the 
physical properties of critical points, nor the experimental constraints discussed in the 
introduction.  Consequently,  the conjectured $\beta$-functions are not well motivated, nor do they
appear to have any reasonable  physical interpretation.

Dolan \cite{D1, D2} considered interpolations between the maximal symmetries 
$\Gamma_X\,(X = R, S, T)$ by postulating a family of meromorphic 
functions that is conjectured to be the $\beta$-function for the RG flow.
However, these ``scaling functions" have no physical foundation or interpretation.
In particular, they have poles where there should be critical points, i.e., where a physical
$\beta$-function must vanish.  Critical exponents are therefore ill defined, and there is no
physical scaling.  

Georgelin et al.\,\cite{GMW} have also considered the symmetry $\Gamma(2)$,
and proposed an interpolating $\beta$-function that in our notation is 
 $\beta\propto (a \lambda - 1)/\theta_3^4$. They conclude (correctly) that the location of 
 the zero of this function is not predicted by $\Gamma(2)$ symmetry, but they seem ambivalent about the order of the transition point, and since the function is holomorphic it cannot be 
 physical.  They do not consider the family of functions under deformations 
 of $a$, and do not discuss the points of enhanced symmetries.

This is in sharp contrast to the idea proposed in Ref. \onlinecite{L1} 
and pursued here.  Our $\beta$-function is derived from a physically motivated and 
physically sensible potential, and therefore by construction is well behaved 
everywhere on the interior of parameter space.  This includes the critical points 
where it has simple zeros, and therefore well defined scaling and critical exponents.

\section{Summary}
\label{sec:summary}
\noindent
A phenomenological analysis using global discrete modular symmetries of RG flows
has previously been successfully applied to the spin-polarized QHE \cite{LR2}.  Following a 
proposal originally made in Ref.\,\onlinecite{L1}, we have here generalized this to 
more complicated situations where spin or other ``flavors" of charge carriers are relevant. 
In some simple cases, including the spin-polarized system, 
maximal subgroups of the modular group 
${\rm PSL}(2,\mathbb Z)$ appear to account for the data, 
while other systems are less symmetric. 

Motivated by the C-theorem, which offers an appealing physical interpretation of the RG potential, 
and insisting on some reasonable physical properties of this potential,  
we are led to propose the simplest possible ansatz for a family of potentials $\Phi_a$ 
that can interpolate between these symmetries (eq.\,(\ref{eq:Phia})).
$\Phi_a$ is parametrized by a single number $a$, up to an undetermined normalization.
The potential is always symmetric under the main congruence group at level two, 
and when $a$ takes certain values this symmetry is enhanced to one of the maximal subgroups 
of the modular group.   The covariant RG $\beta$-function is a holomorphic vector field 
derived from $\Phi_a$, and we have compared the geometry of this gradient 
flow with available temperature driven scaling data. 

Until substantial progress is made in deriving low energy effective field theories
from the micro-physics of quantum Hall systems, $a$ remains an uncomputable, non-universal,
sample specific number that parametrizes our ignorance about these models.  
This is no different from other effective field theories, 
and as usual this is not an obstruction to progress since 
the parameter values can always be simply
extracted from a few experiments if the model is correct.  In our case the value of $a$ is most 
easily determined from experiment by finding the location of a quantum critical point, 
i.e., an unstable zero of the $\beta$-function given by a saddle point of the RG potential.
The surprise is that a single parameter so far has sufficed to account for almost all scaling data, 
in systems with widely different micro-physics but similar emergent modular symmetries.
 
Almost all the data (Refs.\,\onlinecite{LR1,LR2} and Figs.\,6-10) are consistent with 
$a \in \mathbb{R}$, which together with $\Gamma(2)$ symmetry implies a
generalized semicircle law.  The semi-circle law has in some cases been 
confirmed by experiment to be remarkably accurate \cite{semicircle}.

We have argued that this family exhausts all possible RG flows automorphic under the modular group, given some mild constraints on the fixed point structure (the plateaux and quantum critical points) of the global phase diagram of the quantum Hall system. There are two other modular symmetries at level 2, but neither $\Gamma(1)$ nor $\Gamma_P$ have physical potentials. 
Fortunately, there is also no experimental evidence that these symmetries are of interest.

This is one example of a remarkable convergence of quantum Hall physics and modular mathematics.  Hall quantization is, in fact, an automatic and unavoidable consequence 
of modular symmetry.
The geometry of holomorphic modular symmetries just barely admits the existence of functions and forms that we need in order to give a quantitative description of the geometry of observed RG flows, but no more, and unphysical holomorphic $\beta$-functions are prohibited by the symmetry. 

The rigidity of holomorphic modular symmetries allows us to harvest an infinite number of robust predictions: detailed quantitative predictions that are so constraining that it seems unlikely that they can be satisfied by any real physical system. Nevertheless, almost all low temperature quantum Hall scaling data are so far in agreement with these predictions, as shown in 
Ref.\,\onlinecite{LR2} for the spin-polarized case and for the fully spin-degenerate case, 
and for some multi-component systems here.


\end{document}